\documentclass[usegraphicx,usenatbib]{mn2e}

\usepackage{mathptmx}

\newcommand{\Mpc}{\rm\thinspace Mpc}
\newcommand{\kpc}{\rm\thinspace kpc}
\newcommand{\pc}{\rm\thinspace pc}
\newcommand{\km}{\rm\thinspace km}

\newcommand{\cm}{\rm\thinspace cm}

\newcommand{\pcmcu}{\hbox{$\cm^{-3}\,$}}

\newcommand{\yr}{\rm\thinspace yr}
\newcommand{\Gyr}{\rm\thinspace Gyr}
\newcommand{\s}{\rm\thinspace s}

\newcommand{\K}{\rm\thinspace K}

\newcommand{\Msun}{\hbox{$\rm\thinspace M_{\odot}$}}

\newcommand{\Msunpyr}{\hbox{$\Msun\yr^{-1}\,$}}

\newcommand{\keV}{\rm\thinspace keV}

\newcommand{\erg}{\rm\thinspace erg}

\newcommand{\ergpcmsqps}{\hbox{$\erg\cm^{-2}\s^{-1}\,$}}

\newcommand{\ergps}{\hbox{$\erg\s^{-1}\,$}}

\newcommand{\kmps}{\hbox{$\km\s^{-1}\,$}}

\newcommand{\kmpspMpc}{\hbox{$\kmps\Mpc^{-1}$}}

\newcommand{\Zsun}{\hbox{$\thinspace \mathrm{Z}_{\odot}$}}

\newcommand{\psqcm}{\hbox{$\cm^{-2}\,$}}

\begin{document}

\title[Sound waves in the Perseus cluster core]{A deeper X-ray study
  of the core of the Perseus galaxy cluster: the power of sound waves
  and the distribution of metals and cosmic rays}

\author[J.S. Sanders and A.C. Fabian]{J.S. Sanders\thanks{E-mail:
    jss@ast.cam.ac.uk} and
  A.C. Fabian\\
  Institute of Astronomy, Madingley Road, Cambridge. CB3 0HA}
\maketitle

\begin{abstract}
  We make a further study of the very deep \emph{Chandra} observation
  of the X-ray brightest galaxy cluster, A\,426 in Perseus. We examine
  the radial distribution of energy flux inferred by the
  quasi-concentric ripples in surface brightness, assuming they are
  due to sound waves, and show that it is a significant fraction of
  the energy lost by radiative cooling within the inner 75-100~kpc,
  where the cooling time is 4--5~Gyr, respectively. The wave flux
  decreases outward with radius, consistent with energy being
  dissipated. Some newly discovered large ripples beyond 100~kpc, and
  a possible intact bubble at 170~kpc radius, may indicate a larger
  level of activity by the nucleus a few 100~Myr ago.  The
  distribution of metals in the intracluster gas peaks at a radius of
  about 40~kpc and is significantly clumpy on scales of 5~kpc. The
  temperature distribution of the soft X-ray filaments and the hard
  X-ray emission component found within the inner 50~kpc are analysed
  in detail.  The pressure due to the nonthermal electrons,
  responsible for a spectral component interpreted as inverse Compton
  emission, is high within 40~kpc of the centre and boosts the power
  in sound waves there; it drops steeply beyond 40~kpc. We find no
  thermal emission from the radio bubbles; in order for any thermal
  gas to have a filling factor within the bubbles exceeding 50 per
  cent, the temperature of that gas has to exceed 50~keV.
\end{abstract}


\section{Introduction}
The Perseus cluster, Abell~426, is the brightest galaxy cluster in the
sky when viewed in the X-ray band. The cluster contains a bright radio
source 3C~84 \citep{PedlarPerseus90}, the radio lobes of which are
displacing the X-ray emitting thermal gas of the cluster
\citep{BohringerPer93,FabianPer00}. The X-ray emission from the
intracluster medium (ICM) is highly peaked in the centre and the
radiative cooling time of the hot gas is less than 5~Gyr within a
radius of 100~kpc, decreasing to $3\times 10^8\yr$ within the central
10~kpc \citep{SandersPer04}.  A cooling flow of several $100\Msunpyr$
would take place if radiative energy losses from the inner ICM are not
balanced by some form of energy injection. As expected from cooling,
the gas temperature does drop from the outer value of 7~keV within the
central 100~kpc but only down to about 2.5~keV with little X-ray
emitting gas found at lower temperatures, except in coincidence with
line-emitting filaments seen at optical wavelengths
\citep{FabianPer03,FabianPer06}.  Heating by the central radio source
is widely considered responsible for balancing the radiative cooling,
although the exact mechanisms by which the energy is transported and
dissipated and a heating/cooling balance established have been
unclear.

Similar behaviour is found in many X-ray peaked, cool-core clusters
(see \citealt{PetersonFabian06} for a review).  The central galaxy in
such clusters could grow considerably larger than observed if
radiative cooling was unchecked. X-ray observations of their innermost
regions provide an excellent means to study the feedback of an Active
Galactic Nucleus (AGN) on its host galaxy in action. Here we use a
long (900~ks) Chandra observations of the X-ray brightest cluster core
to examine the energy balance and metallicity in considerable detail.
The quality of the data, in terms of counts per arcsec, are much
higher than available with any other cluster. We examine the power
propagating through the core in terms of pressure ripples, or sound
waves, and the cosmic-ray implications of a hard X-ray component. We
also use the metallicity distribution, the temperature profile of the
X-ray emitting gas across an optical filament and assess the thermal
gas content of the radio bubbles.

No X-ray emission has been detected from within the radio bubbles
\citep{SchmidtPer02,SandersPer04}. Two depressions in the X-ray
surface brightness are also found, to the north-west and south, not
associated with high-frequency radio emission. These are likely to be
`ghost' radio bubbles which have detached from the nucleus and
buoyantly risen in the gravitational field
\citep{ChurazovPer00,FabianPer00}, an idea supported by weak low
frequency radio spurs seen pointing towards their direction
\citep{FabianCelottiPer02}.

Surrounding the inner radio lobes are X-ray bright rims
\citep{FabianPer00} at a higher pressure than the outer regions, and
separated from them by a weak isothermal shock \citep{FabianPer06}.
Extending further into the clusters are concentric fluctuations in
surface brightness. These are plausibly pressure and density ripples,
in which case they are sound waves generated by the inflation of the
radio bubbles \citep{FabianPer03,FabianPer06}. The period of the waves
is about 10~Myr, close to the expected age of the bubbles due to
buoyancy, and is not plausibly related to any other source of
disturbance. Ripples due to sound waves have since been found in
simulations \citep{Ruszkowski04,Sijacki06}.  Such sound waves can
transport significant energy in a roughly isotropic manner and so
balance radiative cooling of the intracluster gas if they dissipate
their energy over the surrounding 50--100~kpc
\citep{FabianPer03,Ruszkowski04,FabianReynolds05}.  Concentric
ripple-like features are also found around M87 in the Virgo cluster,
and are interpreted there as weak shocks \citep{FormanM8706}. More
powerful outbursts include those discovered in MS~0735.6+7421
\citep{McNamara05}, Hercules A \citep{NulsenHerc05} and Hydra A
\citep{NulsenHydra05}.

Around the central galaxy in the Perseus cluster, NGC~1275, lies a
giant line-emitting nebula \citep{Lynds1970,Conselice01}, associated
with cool $\sim 0.7$~keV X-ray emitting filaments
\citep{FabianPerFilament03} of $\sim 10^{9}\Msun$ mass
\citep{FabianPer06}. The H$\alpha$ emitting filaments appear to be
have been drawn out of the central galaxy by the rising bubbles
\citep{HatchPer06}.

There is spectral evidence for hard X-ray emission from the central
region, found either with hot thermal \citep{SandersPer04} or
non-thermal \citep{SandersNonTherm05} models. The 2-10~keV luminosity
of this emission is $\sim 5 \times 10^{43} \ergps$.

The iron metallicity structure in the core of the cluster is
inhomogeneous and complex
\citep{SchmidtPer02,SandersPer04,SandersNonTherm05} with the abundance
dropping in the very core. There are also structures such as
high-metallicity ridge and blobs, which may be associated with the
bubbles.

We use a Perseus cluster redshift of 0.0183 here, which gives an
angular scale of 0.37~kpc per arcsec, assuming $H_0 = 70
\kmpspMpc$.

\section{Data preparation}
The datasets analysed in this paper are those that were examined in
\cite{FabianPer06}.  However here we used the standard \textsc{ciao}
data preparation tools on the event files. We used the datasets from
the \emph{Chandra} archive which had gone through Reprocessing III
(reprocessed using version 7.6.7.1 of the pipeline).  Datasets 03209
and 04289 were not reprocessed at the time of the analysis, so we
therefore went through each of the steps in the Science Threads to
reprocess the data manually for these. The reprocessing was done prior
to CALDB 3.3.0 (using gainfile acisD2000-01-29gain\_ctiN0005.fits).

We filtered the level 2 event files for flares as in
\cite{FabianPer06}. We then reprojected each dataset to match the
04952 observation in sky coordinates. We also reprocessed a combined
980-ks blank sky observation file to use the same calibration files as
used by the foreground observations. We randomised the order of the
events in the background file in order to remove any potential
spectral variability.  The background file was split into sections, to
provide a background for each foreground observation.  The length of
each section was chosen to have the same ratio to the total as the
ratio of its respective foreground dataset to the total foreground.
The exposure time of each section was adjusted to give the same count
rate in the 9-12 keV band as its respective foreground (where there is
no source). The background sections were then reprojected to the
original level 2 event file for the matching observation, and then
reprojected to the 04952 observation.

In addition we constructed separate background event files for each
dataset to account for out-of-time events, where photons hit the
detector while it is being read out. We used the
\textsc{make\_readout\_bg} script (written by M.~Markevitch) to
construct these files from the original level 1 event files after
filtering bad time periods. The readout backgrounds were reprojected
to match the 04952 observation.

For the spectral analysis, for a particular region we extracted
foreground spectra from each of the foreground event files relevant
for the region in question. Similarly we made spectra from the
background event files and and readout background files. The
foreground spectra were added together to make a total foreground
spectrum with a total exposure time. Background spectra were added
similarly to make a total background, and so were readout background
spectra.

We also created responses and ancillary responses for each of the
foreground datasets, weighting the responses according to the number
of counts in each spatial region between 0.5 and 7~keV. A response and
ancillary response was made for the total foreground spectrum by
adding the responses and ancillary responses for the individual
observations, weighting according to the number of counts between 0.5
and 7~keV.

To analyse the spectra, we fit them in \textsc{xspec} version 11.3.2
\citep{ArnaudXspec}. The energy range 0.6 to 8~keV was used during
fitting, and the spectra were grouped to have at least 20 counts per
spectral bin. The spectra extracted from the appropriate region from
the background files was used as a background spectrum, and the
spectra from the region from the readout background files was used as
a `correction file'.

In this paper we use the \textsc{apec} \citep{SmithApec01} and
\textsc{mekal}
\citep{MeweMekal85,MeweMekal86,KaastraMekal92,LiedahlMekal95} thermal
spectral models. To model photoelectric absorption we use the
\textsc{phabs} model \citep{BalucinskaChurchPhabs92}.

\section{X-ray surface brightness}

\subsection{Surface brightness images}
\label{sect:sb}
\begin{figure*}
  \includegraphics[width=\textwidth]{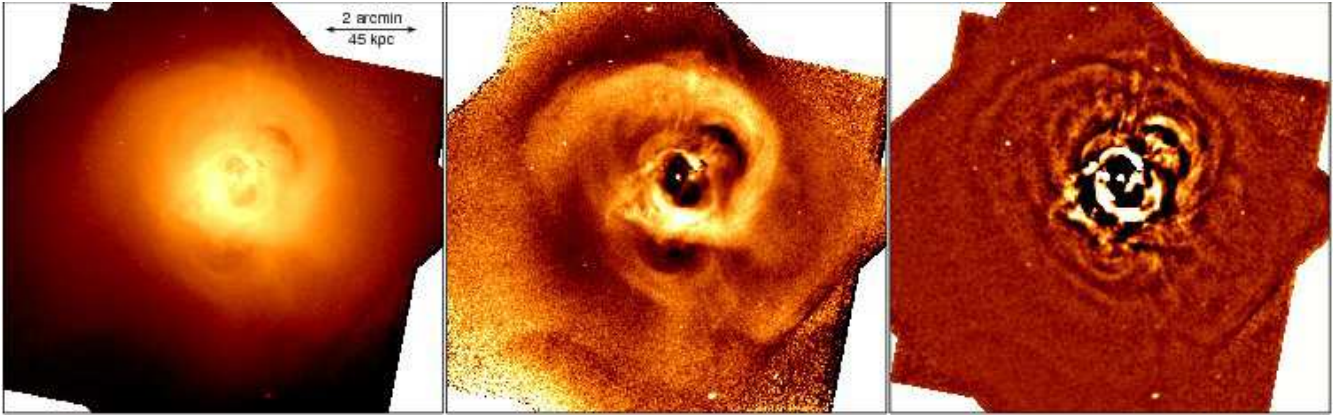}
  \caption{Surface brightness images of the cluster. (Left) 0.3 to 7
    keV full band X-ray exposure-map-corrected image, smoothed with a
    Gaussian of 1.5~arcsec. (Centre) Image after subtracting King
    model fits to 40 sectors, smoothed with a Gaussian of 1.75~arcsec.
    (Right) Original image after high-pass filtering, then smoothing
    with a Gaussian of 1.5~arcsec.}
  \label{fig:sb}
\end{figure*}

\cite{FabianPer03} and \cite{FabianPer06} used unsharp-masking
techniques to reveal the surface brightness fluctuations in the
intracluster medium. Unsharp masking increases the noise in the outer
parts of the image where there are relatively few counts. We have
experimented with several techniques to improve on simple unsharp
masking. We split the exposure-map-corrected image (Fig.~\ref{fig:sb}
left panel) into 40 sectors (centred on the central nucleus), and
fitted a King model to each of the sectors outside of a radius of
13~arcsec (to avoid the central source). We constructed a model
surface brightness image by iterating over each pixel, using the value
obtained by interpolating in angle between the model surface
brightness profiles of the two neighbouring sectors. This model image
was then subtracted from the original image, resulting in
Fig.~\ref{fig:sb} (centre panel). The image very clearly highlights
the surface brightness increase associated with the low-temperature
spiral in the cluster \citep{FabianPer00,ChurazovPer00}. A number of
other features can be seen, including the possible cold front to the
south, the `bay' and the `arc' (figure 2 in \citealt{FabianPer06}). It
does not show the ripples particularly clearly (except perhaps by the
southern bubble) as the cool swirl is dominant.

The ripples are more clearly highlighted using a Fourier high-pass
filter technique. A two-dimensional fast Fourier transform of the
exposure-map-corrected image was made. We removed the low frequency
components with a wavelength greater than 75~arcsec. Frequency
components between wavelengths of 75 and 38~arcsec were allowed
through using with a linear filter increasing from 0 to 1 between
these wavelengths. All shorter wavelengths were left to remain. The
Fourier-transformed image was then transformed back to give
Fig.~\ref{fig:sb} (right panel) after light smoothing. This technique
removes the cool swirl and the underlying cluster emission. It clearly
reveals the ripples, presumably sound waves generated by the inflation
of the bubbles, discovered by \cite{FabianPer03}. It shows a
previously unseen ripple near the edge of the image to the east.

\subsection{Sector across eastern ripples}
\begin{figure}
  \includegraphics[width=\columnwidth]{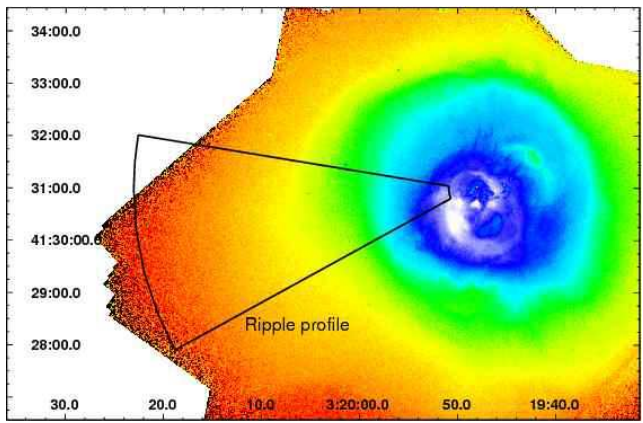}
  \caption{Full band X-ray image showing the sector examined for
    surface brightness fluctuations (Fig.~\ref{fig:sbripple}). The
    region is centred on the inner NE radio bubble to better match the
    surface brightness contours.}
  \label{fig:sbregions}
\end{figure}

\begin{figure}
  \includegraphics[width=\columnwidth]{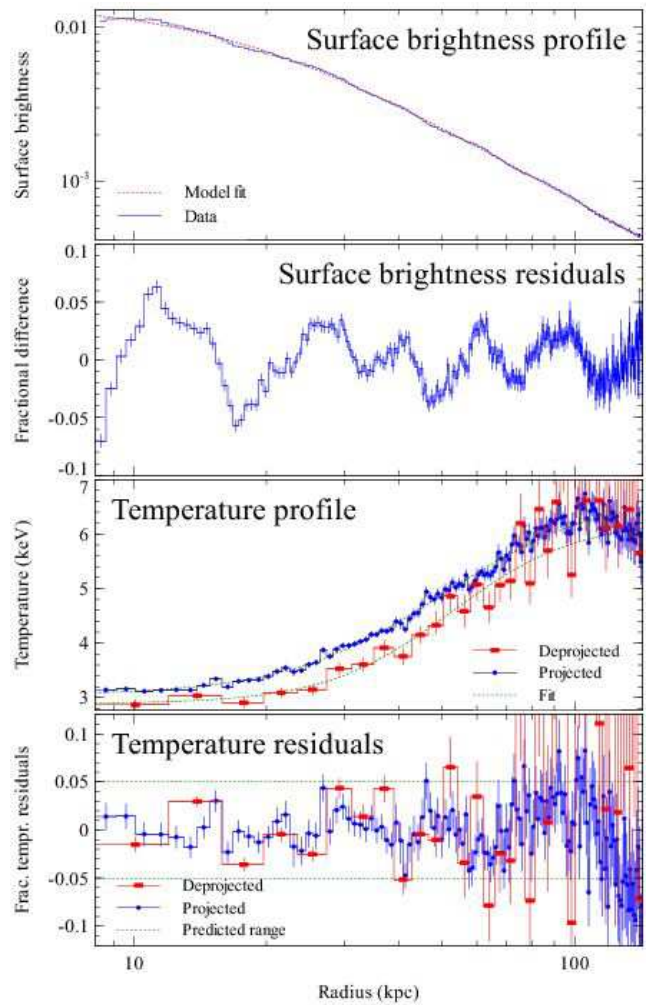}
  \caption{Surface brightness profile to the ESE of the cluster core
    (0.3 to 7 keV), also showing a King model fit, fractional
    residuals from the fit, a temperature profile with fit, and
    fractional residuals from the temperature fit. The dotted lines on
    the temperature residual plot show the expected variation
    associated with a 3 per~cent surface brightness fluctuation. Note
    that the profile is not centred on the nucleus.}
  \label{fig:sbripple}
\end{figure}

To quantitatively examine the ripples, we have examined the surface
brightness in a sector shown in Fig.~\ref{fig:sbregions}. We show in
Fig.~\ref{fig:sbripple} (top panel) a surface brightness profile (in
the 0.3 to 7~keV band) made to the east-south-east (ESE) of the
cluster core. Also shown in the top panel is a King model fit to the
profile. In the second panel we show the fractional residuals to the
fit, clearly showing the oscillations in surface brightness observed
in Fig.~\ref{fig:sb} (right panel) at the few percent level (the first
peak in this figure corresponds to the bright rim of the radio
bubbles).

To look for any temperature structure associated with the oscillations
we fitted \textsc{phabs} absorbed \textsc{apec} thermal spectral
models to spectra extracted in annuli from the spectra. The fitting
procedure allowed the temperature, metallicity, normalisation and
absorption to vary, and minimised the C-statistic \citep{Cash79} in
each fit. The projected temperature profile is shown in the third
panel of Fig.~\ref{fig:sbripple}. We also show the results from
fitting the same model (using the $\chi^2$ statistic) to deprojected
spectra in larger bins. We used the deprojection method in
Appendix~\ref{appendix:deproj} to create the deprojected spectra.

We finally show the residuals from a simple `$\eta$ model'
\citep{AllenSchmidtFabian01} fits to the projected and deprojected
temperature profiles (Fig.~\ref{fig:sbripple} fourth panel)
superficially.  There is no obvious temperature structure associated
with the ripples in this location of the cluster.  The amount of
temperature variation expected from the surface brightness
fluctuations can be estimated.  Simulations of thermal spectra in
\textsc{xspec} shows that the surface brightness is independent of
temperature at constant density in the temperature range 3--6~keV. If
the adiabatic index of the ICM is $\gamma=5/3$, assuming the ideal gas
law and that the X-ray surface brightness is proportional to the
density squared, the fractional temperature ($T$) fluctuations
associated with surface brightness ($I$) changes should be of
magnitude
\begin{equation}
  \frac{\delta T}{T} = \frac{1}{3} \frac{\delta I}{I}.
\end{equation}
We plot on the lower panel of Fig.~\ref{fig:sbripple} dotted lines
showing the range of temperature variation expected to be associated
with 3~per~cent variations in surface brightness (which are the
maximum observed here). These include a scaling factor of 5 to convert
from projected surface brightness fluctuations to intrinsic emissivity
fluctuations (see Section \ref{sect:wavepower}). The deprojected
temperatures are comparable to those expected but we caution that
there is significant noise in the results.

\subsection{Wave power}
\label{sect:wavepower}
We now estimate the power implied by the surface brightness
fluctuations observed in the cluster (Fig.~\ref{fig:sb} right),
assuming that they are sound waves. This is then compared with the
power radiated within the inner regions of the cluster core where the
radiative cooling time is shorter than its expected age of a few Gyr
(by age in this context we mean the time since the last major merger).

The instantaneous power, $\mathcal P$, transmitted in a spherical
sound wave is given by \citep{LandauLifshitzFluids}
\begin{equation}
  \mathcal P = 4 \pi r^2 \frac{\delta P^2}{\rho c},
  \label{eqn:wavepower}
\end{equation}
at a radius $r$, where the sound wave pressure amplitude is $\delta
P$, the mass density of the medium is $\rho$ and the sound speed is
$c$. The sound wave pressure amplitude can be computed from the
density amplitude with
\begin{equation}
  \delta P = \frac{5}{3} \, n_\mathrm{e} \, kT \,
  \frac{\delta n_\mathrm{e}}{n_\mathrm{e}}\alpha,
\end{equation}
assuming $\gamma=5/3$, where $\alpha$ is a factor to convert from the
electron number density $n_\mathrm{e}$ to the total number density.

The fractional variation in density over the wave $\delta
n_\mathrm{e}/n_\mathrm{e}$ is estimated from the surface brightness
fractional change.  As discussed above, there is little variation of
surface brightness with temperature at constant density in this
temperature range. Assuming bremsstrahlung emission, the fractional
variation in density should be half that of the surface brightness
seen. This is not the case in reality, as projection effects are
dominant. We constructed simple numerical simulations of 10 to 20~kpc
wavelength density waves in a cluster density profile (we tried the
profiles in \citealt{ChurazovPer03} and \citealt{SandersPer04}).  The
typical conversion factor from a fractional surface brightness to
density perturbation is 2--3, with more suppression for smaller
wavelength waves.

We take the surface brightness image, and filter it with a high-pass
filter as in Section~\ref{sect:sb}. We use a looser filtering here as
some of the observed 20~kpc waves are otherwise suppressed (not
filtering anything below a wavelength 62~arcsec, increasingly
filtering up to 124~arcsec and discarding everything longer
wavelength).  The original surface brightness image was binned using
the contour binning algorithm \citep{SandersBin06} to have $10^4$
counts per region. We applied the same binning to the filtered image,
and divided it by the binned surface brightness image.  This created a
fractional surface brightness variation map. The contour-binning
routine follows the surface brightness closely, so bins are also
aligned well with the ripples.

\begin{figure}
  \includegraphics[width=\columnwidth]{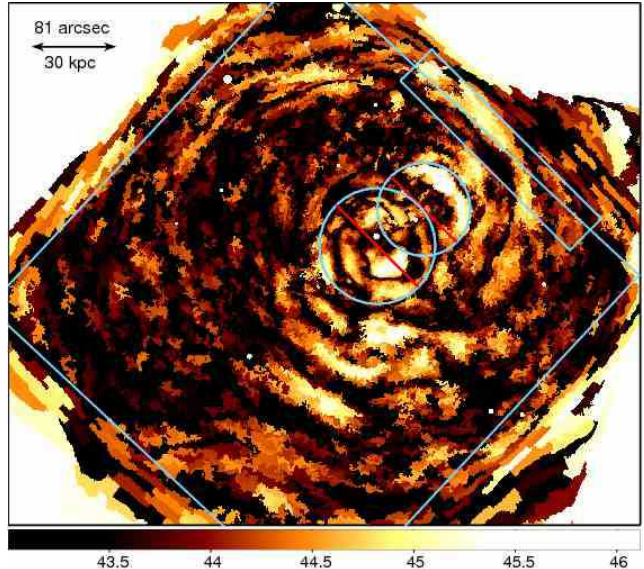}
  \caption{Power in surface brightness fluctuations, assuming they are
    due to spherical waves. The units are $\mathrm{log}_{10}
    \ergps$. The rectangles show the regions examined in
    Fig.~\ref{fig:wavepowcuml} and \ref{fig:wavepowinst}. The circles
    show the excluded inner bubble and NW bubble regions. We assume a
    factor of 2.5 to convert from surface brightness to density
    variations.}
  \label{fig:wavepowerimg}
\end{figure}

For each pixel on the fractional surface brightness variation map, we
compute the power in a spherical wave from Equation
\ref{eqn:wavepower}, assuming that the radius of the wave is the
projected radius on the sky. Deprojected density and temperature
values at each radius were calculated from a fit to the average
profiles in \cite{SandersPer04}. In Fig.~\ref{fig:wavepowerimg} we
show a map of computed power values for each pixel, assuming a factor
of 2 to convert from surface brightness variations to density
variations. Many of the ripple features in Fig.~\ref{fig:sb} (right
panel) are seen in this image, plus the radio bubbles (which are not
themselves sound waves). We bin the data in this image rather than use
simple smoothing, as noise in the outer regions means that the power
is overestimated.

\begin{figure}
  \includegraphics[width=\columnwidth]{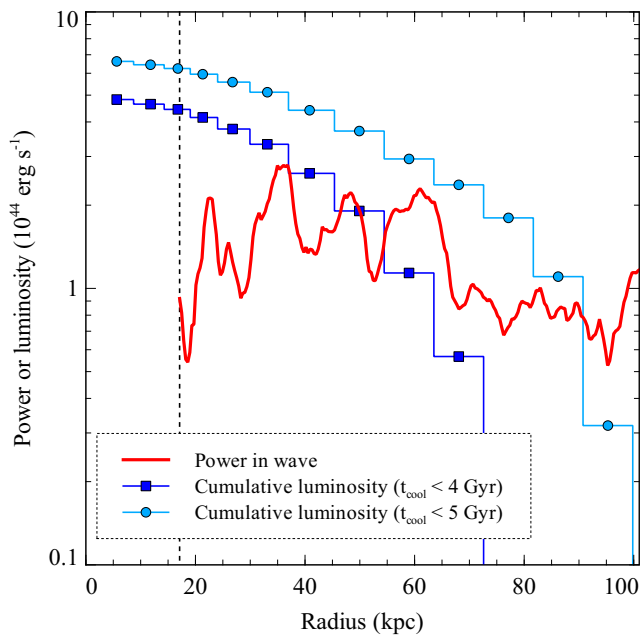}
  \caption{Plot of mean power in sound waves and the cluster
    cumulative X-ray luminosity inwards of where the cooling time is 4
    or 5~Gyr.  The vertical line shows the inner radius of our
    measurements due to the radio lobes. The total power in the lobes
    approaches $10^{45}\ergps$ \citep{DunnFabian04}.}
  \label{fig:wavepowcuml}
\end{figure}

\begin{figure*}
  \includegraphics[width=0.7\textwidth]{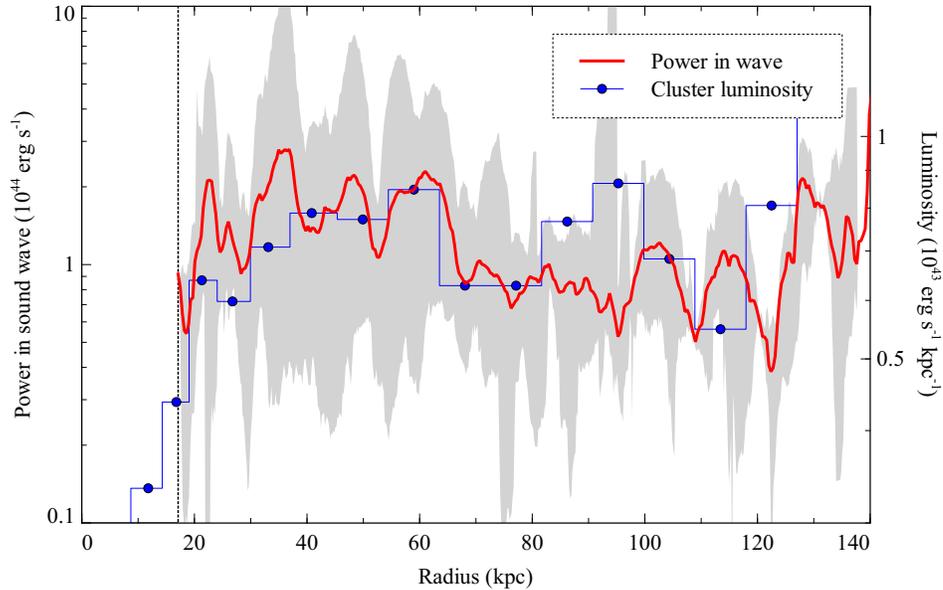}
  \caption{Plot of mean power in sound wave and the cluster luminosity
    per unit length (in radius). The shaded region shows the the
    minimum and maximum sound wave power when the data are split into
    six equal sectors. The cluster luminosity shows the heating rate
    at each radius required to combat cooling. The vertical line shows
    the inner radius of our measurements due to the radio lobes. Note
    the difference in scales, the power being on the left in units of
    $10^{44} \ergps$ and the luminosity on the right in units of
    $10^{43} \ergps \kpc^{-1}$.}
  \label{fig:wavepowinst}
\end{figure*}

In Fig.~\ref{fig:wavepowcuml} we show the average values at each
radius, masking out the central bubbles and north-west bubbles and the
edges of the CCD to avoid filtering artifacts (using the regions in
Fig.~\ref{fig:wavepowerimg}). On the plot we also display the
cumulative luminosity from the cluster calculated from the deprojected
density, temperature and abundance values from \cite{SandersPer04}. We
accumulate the luminosity inwards from a radius of 75 or 100~kpc, the
radii corresponding to where the mean radiative cooling time of the
gas \citep{SandersPer04} is $\sim 4$ or $5 \Gyr$, the likely age of
the cluster.  Note that if the sound speed is a function of azimuth in
the cluster (the temperature map indicates that this is so), then the
phasing of the waves depends on azimuth.  This will lead to smearing
of the ripples in Fig.~\ref{fig:wavepowcuml}.

We plot the wave power out to larger radii in
Fig.~\ref{fig:wavepowinst}. In this graph we also show the variation
in wave power at each radius as a shaded region. This was computed by
repeating the calculation of the average power in six equal sectors,
and shading the region between the minimum and maximum values at each
radius. At large radii only a couple of sectors were used, due to the
position of the source on the detector. We show the luminosity of the
cluster per unit length of radius in this plot to compare to the wave
power.

Fig.~\ref{fig:wavepowcuml} shows that the net power in the ripples is a
few times $10^{44}\ergps$ and sufficient to offset a significant part
of the radiative cooling within the innermost 70~kpc or so. The power
implied by the analysis drops off with radius out to 120~kpc, with an
e-folding length of about 50~kpc, consistent with models of viscous
dissipation \citep{FabianPer03}, which is required if heating by sound
waves offsets radiative cooling.

Larger power is seen near the edge of Fig.~\ref{fig:wavepowinst}
around 105, 115 and 125 kpc radius. These show possible evidence that
the source was more powerful several $10^8\yr$ ago. Such powerful
shocks could have been created as several individual bursts, or from a
single burst producing multiple sound waves \citep{Brueggen07}.  We
caution that most of the signal comes from a small angular region to
the extreme SW of the main detector. Further observations covering a
wider region are required to accurately determine the wave power at
this radius.

\begin{figure}
  \includegraphics[width=\columnwidth]{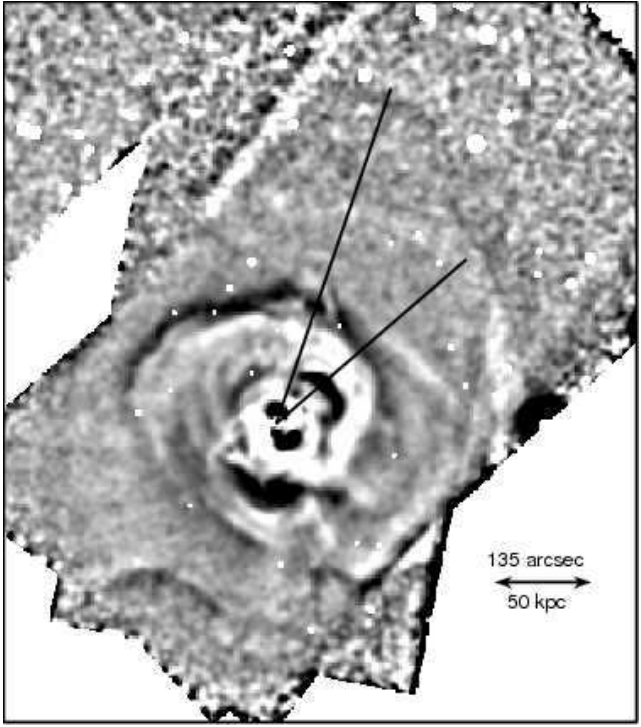}
  \caption{Further possible surface brightness discontinuities to the
    north of the cluster. This is an unsharp-masked image, subtracting
    images smoothed with a Gaussian by 4 and 16~arcsec and dividing by
    the 16~arcsec map. Point sources were excluded from the
    smoothing. The apparent features are at radii of $\sim 130$ and
    170~kpc (indicated by solid lines).}
  \label{fig:unsharp_outer}
\end{figure}

\begin{figure}
  \includegraphics[width=\columnwidth]{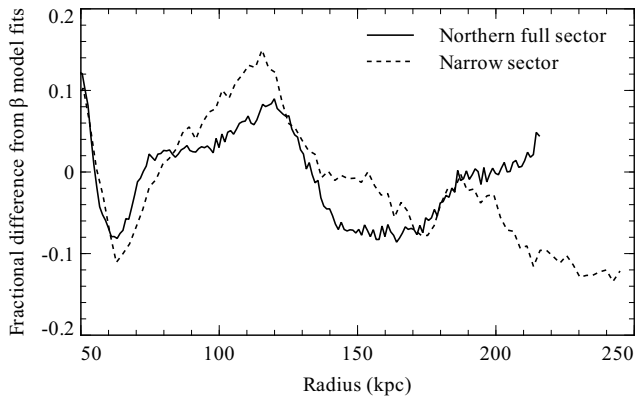}
  \caption{Fractional residuals from a $\beta$ model fit to the
    surface brightness in sectors to the north at large radii. The
    full sector shows a profile between 19 and $75^\circ$ from west
    towards north, while the narrow sector is between 59 and
    $75^\circ$.}
  \label{fig:betafit_outer}
\end{figure}

We also looked for structure in images of the cluster combining all of
the ACIS (Advanced CCD Imaging Spectrometer)
CCDs. Fig.~\ref{fig:unsharp_outer} shows an unsharp-masked image of
the north of the cluster to large radii. We note that there appears to
be a sharp edge at around 130~kpc radius in this direction (although
it varies in radius in the northern sector). This is at approximately
the same radius as the ripple seen in Fig.~\ref{fig:wavepowinst}. The
edge can be seen in residuals from $\beta$ model plus background fits
to a wide and narrow sector (Fig.~\ref{fig:betafit_outer}). This
discontinuity can also be seen in an \emph{XMM-Newton} observation of
Perseus (Fig.~7 in \citealt{ChurazovPer03}).

At a radius of around 170~kpc is another feature. This appears to be a
dip in surface brightness followed by a rise. This feature is
particularly sharp in fits to the surface brightness in a narrow
sector (Fig.~\ref{fig:betafit_outer}). A possible interpretation is an
ancient radio bubble which is still intact. The thermal gas pressure
is likely to be around 4 times less at this radius, so if the bubble
remains intact it will be around 4 times larger than it was originally
(if it retains its original energy). Rough scaling with the rising
bubble NW of the nucleus suggests that it would have to have at least
twice as much energy. The direction from the centre in which the dip
is most visible (the longer solid line in
Fig.~\ref{fig:unsharp_outer}) is also directly along the northern
H$\alpha$ filament and fountain \citep{FabianPer06}. If it is indeed a
bubble then it shows that bubbles remain intact to very large radii in
galaxy clusters. Bubble-like low pressure regions were also seen to
the south of the core \citep{FabianPer06}.

To constrain better the power in the feature at 130~kpc and confirm
the radio bubble near 170~kpc radius, requires further deep
observations by \emph{Chandra} offset from the cluster core to improve
the point spread function (the average radius of which is 10~arcsec at
10~arcmin off axis, compared with $\sim 1$~arcsec on axis).

\section{Metallicity map}
\begin{figure}
  \includegraphics[width=\columnwidth]{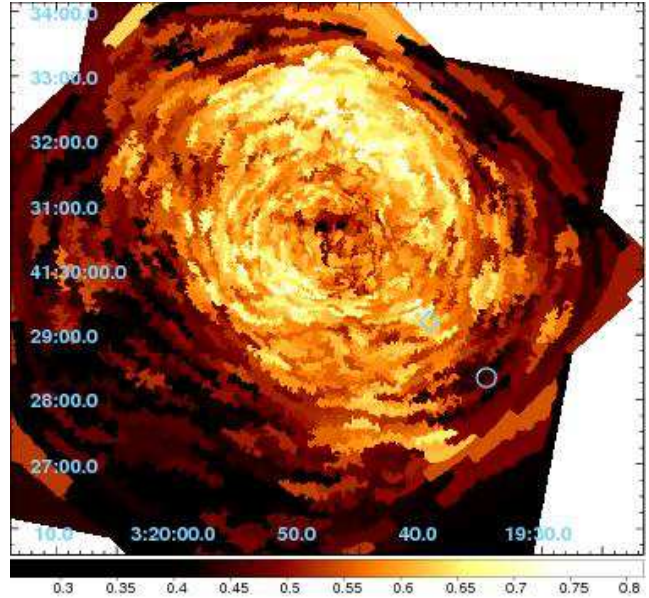}
  \caption{Metallicity map of the core of the cluster relative to
    solar. Regions contain greater than $4\times 10^{4}$ counts. Fits
    assume solar ratios of elements, but the results mainly depend on
    the iron abundance. Uncertainties for each spectral fit range
    smoothly from $0.025$ in the centre to $0.07\Zsun$ to the extreme
    bottom left of this image. The circle shows the approximate
    position of the ancient bubble, and the diamond shows the high
    metallicity blob in its apparent wake.}
  \label{fig:Zmap}
\end{figure}

Fig.~\ref{fig:Zmap} shows a metallicity map of the core of the
cluster. This was generated by extracting spectra from contour-binned
regions \citep{SandersBin06} containing $\sim 4 \times 10^4$ counts.
The spectra were fit by a \textsc{phabs} absorbed \textsc{mekal} model
with the absorption, temperature, emission-measure and metallicity
free. Note that the plot does not clearly show the high abundance
shell (possibly marking the location of an ancient bubble) found by
\cite{SandersNonTherm05} as there were no new data in that region in
these observations, and the spatial regions we use here are larger.

\begin{figure}
  \includegraphics[width=\columnwidth]{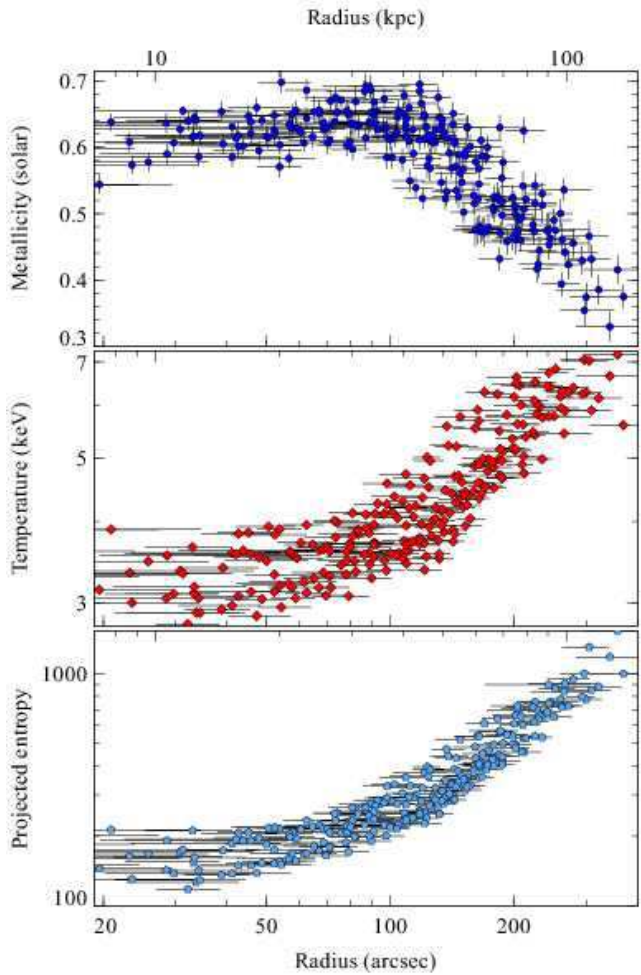}
  \caption{Radius versus projected temperature and metallicity. Values
    were measured from spectral fitting a \textsc{mekal} model to
    spectra from bins with $\sim 2.5 \times 10^5$ counts. Also plotted
    is the projected entropy as a function of radius (see Section
    \ref{sect:metalrelations}).}
  \label{fig:rad_Z_T}
\end{figure}

To examine the variation more quantitatively, we have repeated the
spectral fitting using regions containing $\sim 2.5 \times 10^5$
counts to decrease the size of the uncertainties.  The radial
temperature and metallicity variation are plotted in
Fig.~\ref{fig:rad_Z_T}, generated by plotting the average radius of
each bin against the value obtained from that region. At each radius
there is a large spread in temperature and abundance.

\subsection{Metallicity relations}
\label{sect:metalrelations}
\begin{figure}
  \includegraphics[width=\columnwidth]{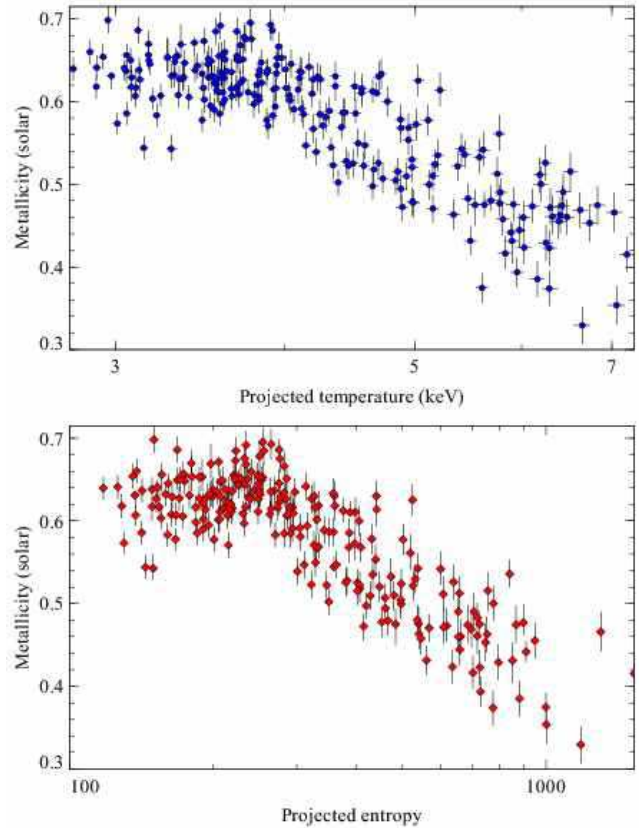}
  \caption{Temperature versus metallicity and pseudo-entropy versus
    metallicity plots. All values are projected.  The bins used are
    the same as in Fig.~\ref{fig:rad_Z_T}.}
  \label{fig:Z_entropy_T}
\end{figure}

In \cite{SandersPer04} we plotted the temperature of regions against
their metallicity. As the temperature of the gas declines towards the
centre, the metallicity reaches a maximum at a radius of 40~kpc, then
decreases again. We plot the metallicity-temperature relation from the
new data in Fig.~\ref{fig:Z_entropy_T} (top panel).  Although we see a
similar relation, the reduced errors bars show that there is
significant metallicity scatter at each temperature. Indeed the
scatter appears similar to that in the radial plot of the metallicity
(Fig.~\ref{fig:rad_Z_T}). The temperature of the gas does not appear
to correlate better with the metallicity than the radius.

Another interesting physical quantity is the entropy of the gas.
Entropy in clusters is usually defined as $K = kT n_{e}^{-2/3}$.
Using the \textsc{xspec} normalisation\footnote{\textsc{xspec}
  \textsc{mekal} and \textsc{apec} normalisations are defined as
  $\{10^{-14} \int n_\mathrm{e} n_\mathrm{H} \mathrm{d}V\} / \{4 \pi
  [D_A (1+z)]^2\}$, where the source is at redshift $z$ and angular
  diameter distance $D_A$, and the electron number density
  $n_\mathrm{e}$ and Hydrogen number density $n_\mathrm{H}$ are
  integrated over volume $V$.} per unit area on the sky, $N \propto
n_{e}^{2} d$, where $d$ is a depth in the cluster, which we assume to
be constant, we can calculate a pseduo-projected entropy quantity
$N^{-1/3} kT$.  The plot of this quantity against the metallicity
(Fig.~\ref{fig:Z_entropy_T} bottom panel) looks similar to the
temperature-metallicity plot, with a great deal of scatter at each
entropy value. The values use $N$ in \textsc{xspec} normalisation
units per square arcsec and $kT$ in keV.

The scatter in the abundance seems unrelated to the temperature,
radius, or entropy. Higher metallicity regions will have shorter mean
radiative cooling times as the line emission is stronger, but this
effect is not strong at temperatures of $\sim 3-7 \keV$, so it is
perhaps not surprising that metallicity and temperature or entropy
(and therefore cooling time) are unrelated.

\subsection{Central abundance drop}
There is also a drop in metallicity in the central regions (as can be
seen in the metallicity map in Fig.~\ref{fig:Zmap}). This feature
appears to be unaffected by including other components, such as extra
temperature components, or powerlaw models.

\begin{figure}
  \includegraphics[width=\columnwidth]{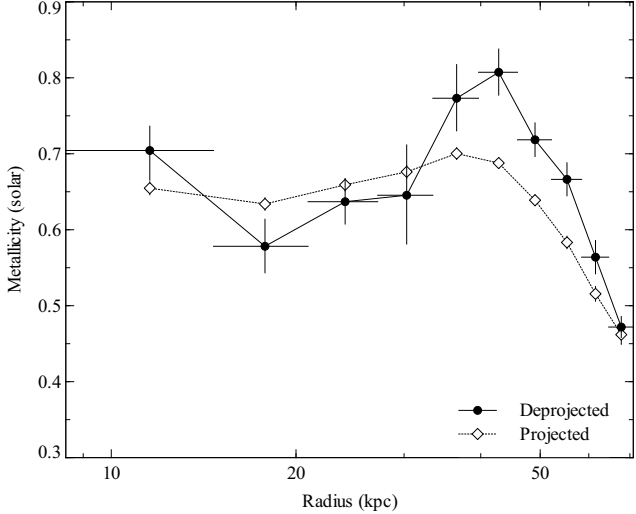}
  \caption{Iron metallicity profile to the NW of the cluster between
    position angles 290.0 and $4.7^\circ$ from 3C~84. This was
    produced by fitting \textsc{vmekal} models to projected and
    deprojected spectra.}
  \label{fig:Z_deproj}
\end{figure}

We have investigated whether the central drop in metallicity could be
caused by projection effects. We use the spectral deprojection method
outlined in Appendix~\ref{appendix:deproj} to construct a set of
deprojected spectra. Fig.~\ref{fig:Z_deproj} shows the iron
metallicity profile computed by fitting isothermal \textsc{vmekal}
models to the spectra before and after deprojection.  In the fit the
elemental abundances of O, Ne, Mg, Si, S, Ar, Ca, Fe and Ni were
allowed to vary, with the gas temperature, the model emission measure,
and the absorbing column density. The analysis shows the peak at
around 40~kpc radius is enhanced by accounting for projection and the
central drop remains. 

The drop in metallicity in the central regions appears to be robust.
The entropy of the gas in the central regions is much lower than
further out, which means it is difficult for the source of the low
metallicity gas to be from larger radius.

\subsection{Small scale variation}
\begin{figure}
  \includegraphics[width=\columnwidth]{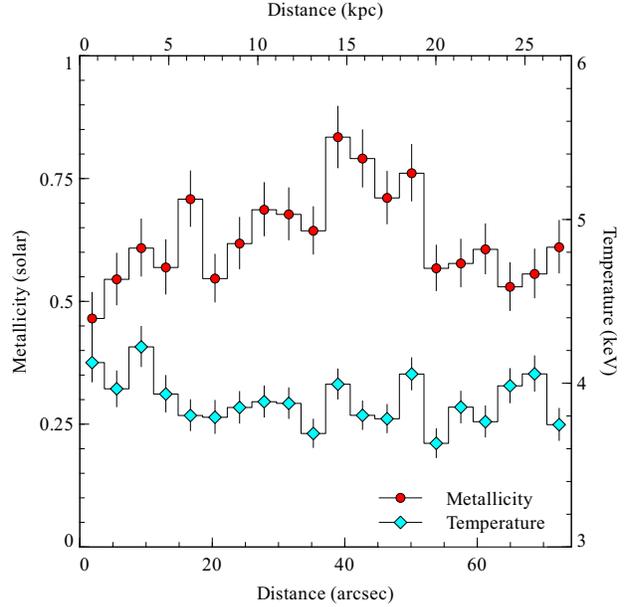}
  \caption{Abundance and temperature profiles across the high
    metallicity blob near 3\fh19\fm40\fs, $+41^\circ29'15''$. The
    profile crosses the blob in the SE to NW direction, inclined by an
    angle of $40^\circ$ from the west northwards.}
  \label{fig:blob_profile}
\end{figure}

Much of the structure in the metallicity map is real. To demonstrate
this, we show in Fig.~\ref{fig:blob_profile} a metallicity and
temperature profile across the high metallicity blob apparently in the
wake of a possible ancient bubble down to the extreme SW (marked on
Fig.~\ref{fig:Zmap}; \citealt{SandersNonTherm05}). The metallicity
profile shows a significant peak with a width of $\sim 5 \kpc$ at the
position of the blob (we also see $\sim 5 \kpc$ metallicity features
near the filaments in Section~\ref{sect:proffilament}). There is no
obvious correlation between the temperature and metallicity
profiles. Taking the global diffusion coefficient of $2 \times 10^{29}
\cm^{2} \s^{-1}$ measured by \cite{RebuscoDiff05} in Perseus and a
lengthscale of 5 kpc, the lifetime of such a feature is only $\sim
40$~Myr. This is roughly comparable to the buoyancy timescale
estimated for the ancient bubble of 100~Myr \citep{DunnPer06}, given
the large uncertainties, but may require that diffusion is suppressed
on small scales as it appears unlikely that the metals could have been
injected in-situ. The sharp edges of the feature would imply diffusion
times of only a few Myr, requiring significant suppression.

The blob has a metallicity approximately $0.2 \Zsun$ higher than its
surroundings (which are at approximately $0.6 \Zsun$). This is
probably a lower limit because of projection effects. Assuming the
blob is sphere with diameter 5~kpc, and a local electron density of
around $0.037 \pcmcu$ \citep{SandersPer04}, it represents an
enhancement of around $2.6 \times 10^4 \Msun$ of Fe. If most of this
enrichment is due to Type~Ia supernova, then this corresponds to
around $3.7 \times 10^4$ supernova (assuming $0.7 \Msun$ of Fe
produced per supernova). Taking the timescale from diffusion, this
corresponds to 0.1 supernova type Ia per century for this 5~kpc radius
region, which is a significant fraction of the rate expected from a
single galaxy.

\section{The high velocity system}
\label{sect:hvs}
\cite{GillmonPer04} studied the High Velocity System (HVS), a distinct
emission-line system at a higher velocity than NGC~1275
\citep{Minkowski57}, using an earlier 200-ks \emph{Chandra}
observation of the system. The system is observed in absorption in
X-rays \citep{FabianPer00}, as it lies between most or all of the
cluster and the observer. \cite{GillmonPer04} mapped the absorbing
column density and placed a lower limit of 57~kpc of the distance
between the nucleus of NGC~1275 and the HVS. They obtained a total
absorbing gas mass of $1.3 \times 10^9 \Msun$, assuming Solar
metallicities.

\begin{figure}
  \includegraphics[width=\columnwidth]{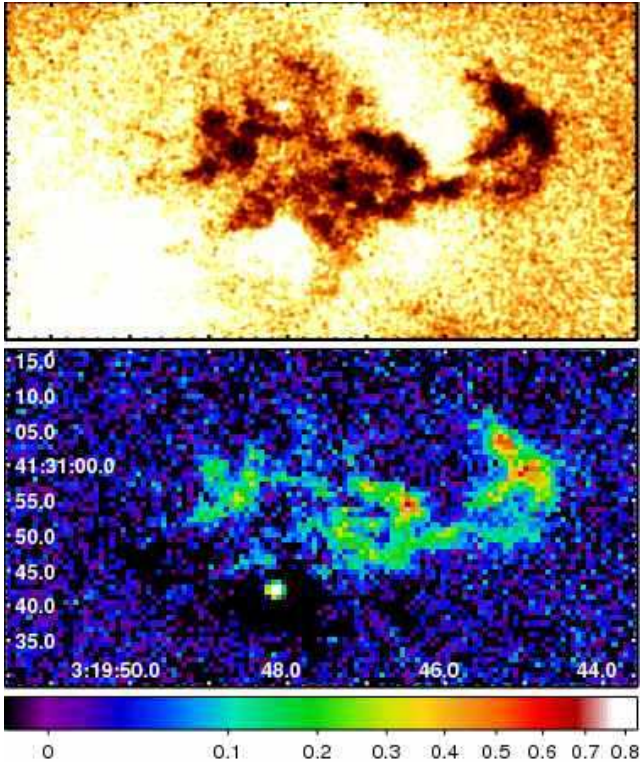}
  \caption{(Top panel) Image of the high-velocity system in the 0.3 to
    0.7 keV band smoothed by a Gaussian of 0.49~arcsec. (Bottom panel)
    Measurements of the equivalent Hydrogen column density over the
    high-velocity system region, in units of $10^{22}\psqcm$
    (subtracting Galactic absorption of $0.12 \times 10^{22}\psqcm$).
    The values were measured from spectral fitting in $0.74 \times
    0.74$ arcsec pixels. The central nucleus of 3C~84 shows up as a
    point-like object near 3\fh19\fm48\fs, $+41^\circ30'40''$.  The
    uncertainty on the column density on individual pixels is around
    0.1 for the regions of highest absorption and 0.03 outside of the
    HVS.}
  \label{fig:NHmap}
\end{figure}

Here we repeat the mapping analysis with the new 900-ks dataset.
Fig.~\ref{fig:NHmap} (top panel) shows an image in the 0.3-0.7~keV
band, clearly showing the absorbing material. To quantify the amount
of material we fitted an \textsc{apec} thermal spectral model for the
cluster emission, absorbed by a \textsc{phabs} photoelectric absorber
to spectra extracted from a grid of $0.74 \times 0.74$ arcsec regions
around the HVS. In the fits the temperature of the thermal emission,
its normalisation, and the absorbing column density were free. The
abundance of the thermal gas was fixed to $0.7 \Zsun$ (based on the
results of spectral fitting to larger regions near the core of the
cluster; the results are very similar if this is allowed to be free).
Spectra were binned to have a minimum of 5 counts per spectral
channel. We minimised the C-statistic to find the best fitting
parameters (note that no background or correction spectra were used in
the fits here, as the cluster is much brighter than the background
here).

In Fig.~\ref{fig:NHmap} (bottom panel) we show the resulting Hydrogen
column density map over the HVS region, including the Galactic
contribution. We stress that these values are \emph{equivalent}
Hydrogen column density. The measurements are most sensitive to the
abundance of Oxygen in the absorber. The Hydrogen column density is
calculated assuming the Solar abundance ratios of
\cite{AndersGrevesse89}, which gives an Oxygen to Hydrogen number
density ratio of $8.51 \times 10^{-4}$. The column density is flat in
this region beyond the HVS.  We obtain an average value of this
Galactic component of $0.12 \times 10^{22} \psqcm$, consistently from
several surrounding regions.

The total number of absorbing atoms can be calculated using the mean
absorption over the HVS, the Galactic contribution, and the distance
to Perseus. We converted this into a total absorbing mass of $1.1
\times 10^9 \Msun$, assuming Solar metallicity ratios. This is lower
than the value of $(1.32 \pm 0.05) \times 10^9 \Msun$ quoted by
\citeauthor{GillmonPer04}. We repeated our analysis using the
1.96~arcsec binning factor used in that paper, but this did not
significantly change our result. The twenty per cent difference may be
due to the different calibration used in the earlier analysis,
particularly the lack of correction for the variation in the
contaminant on the detector, as the Galactic value is important to
determine the total mass.

To place an improved lower limit of the distance of the HVS from the
cluster nucleus, we examined a 0.3 to 0.7 keV image, where the
absorption is strongest. Taking 1-1.5~arcsec diameter regions (with
total area 3.9 arcsec$^2$) in the three highest absorption regions we
find 17 counts in total in this band. Exposure-map correcting this
surface brightness, and comparing it to the exposure-map corrected
image of the cluster, we find that the surface brightness from the
cluster only goes down to this level at a radius of approximately
110~kpc. This is a lower limit of the distance of the HVS from the
cluster centre, as if the HVS were closer, the cluster emission should
be `filling in' the decrement in count rate caused by the absorption
\citep{GillmonPer04}.  This lower limit is a significantly better the
previous value of 57~kpc.

As we discuss in Section \ref{sect:hvsthermal}, the HVS could have a
shock cone behind it if it is travelling exactly along our line of
sight towards the cluster. If this is the case, the shocked material
would have a significantly lower density than the cluster at the same
radius. This would mean that the lower limit we compute above would be
overestimated.

\section{Profiles across filaments}
\label{sect:proffilament}
\begin{figure}
  \includegraphics[width=\columnwidth]{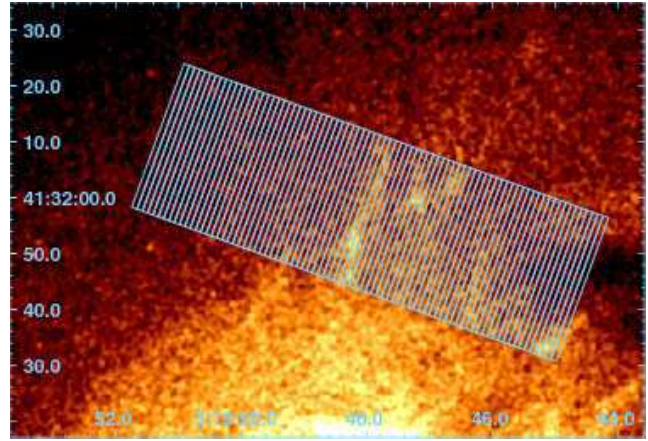}
  \caption{Location of the regions used to measure the thermal
    properties across the filaments, displayed on a 0.3 to 1 keV
    image. Each of the 80 regions is $1\times27.3$ arcsec in size.}
  \label{fig:filamentregions}
\end{figure}

To investigate the thermal structure of the X-ray filaments associated
with the H$\alpha$ nebula in detail, we extracted spectra from small
$1 \times 27.3$~arcsec boxes in a profile across the filament (the
regions we used are shown in Fig.~\ref{fig:filamentregions}). We
created background spectra, responses, ancillary responses, and out of
time background spectra for each region.

\begin{figure}
  \includegraphics[width=\columnwidth]{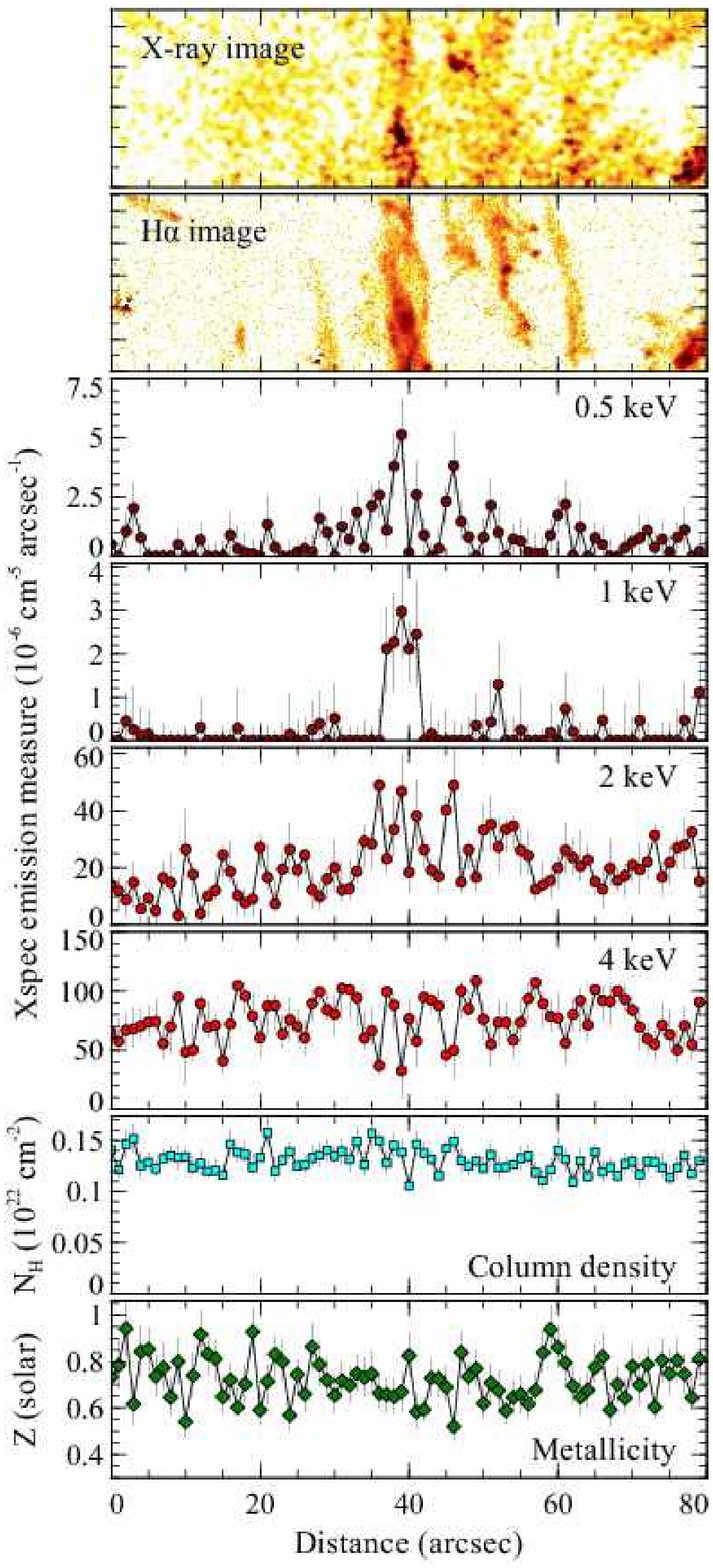}
  \caption{Emission measure profiles across the filaments in the
    different temperature components. The top panel shows an
    unsharp-masked 0.5-7~keV X-ray image rotated so that the bins lie
    across it. The second panel shows a similar H$\alpha$ image. The
    next panels show the 0.5, 1, 2 and 4~keV temperature component
    \textsc{xspec} normalisations, measured from the 1~arcsec wide
    bins. The final panels show the best fitting absorbing Hydrogen
    column density and the metallicity of thermal components.}
  \label{fig:filamentmulti}
\end{figure}

\subsection{Multiphase model}
Our first model was to fit the spectra with a multiphase model
consisting of \textsc{apec} thermal components at fixed temperatures
of 0.5, 1, 2, 4 and 8~keV. The normalisations were allowed to vary and
the metallicities of each of the components were tied together. They
were absorbed with a \textsc{phabs} photoelectric absorber which was
allowed to vary. We assume each component has the same metallicity as
we cannot measure them independently. The measurement is likely to be
driven by the cooler components as they are line-dominated.  The model
is similar to that used to produce figure 12 in \cite{FabianPer06},
mapping the multiphase gas. We show the emission measure of each
temperature component in Fig.~\ref{fig:filamentmulti}, with the
absorbing column density and metallicity. Also shown is an
unsharp-masked image of the region examined in the 0.5-7~keV X-ray
band (the X-ray image has been rotated so that the 1-arcsec bins lie
horizontally across the region) and the H$\alpha$ image from
\cite{Conselice01} taken using the WIYN telescope. We chose the
rotation angle so that the bins were lined up across the central X-ray
features. The H$\alpha$ filaments are not quite aligned in angle.

The plot shows that the filaments do not contain gas at temperatures
of $\sim 4$~keV. At 2~keV we start to see the filaments, although
their signal is fairly weak against that from nearby gas. The
filaments are very strong near 1~keV, and still visible at 0.5~keV.
There are some point-to-point differences however. The strong filament
at 39~arcsec distance is strong in 1 and 0.5 keV, but the one nearby
46~arcsec does not appear at 1~keV, but does at 2 and 0.5 keV.

The column density profile is fairly flat. A linear model fitted to
the column density profile gives a reduced $\chi^2$ of
$86/78=1.10$. We see no evidence for additional X-ray absorption
associated with the filaments.

The metallicity profile shows several features, which matches the
complex structure seen in the metallicity map (Fig.~\ref{fig:Zmap}). A
simple linear fit gives a reduced $\chi^2$ of $105/78=1.34$. There are
quite strong peaks at the 59~arcsec, and at 12~arcsec. Neither is
exactly at the position of a filament, though the one at 59~arcsec is
offset by an arcsec or two from some gas from 2 to 0.5~keV. It is
possible that some of the metallicity variation is caused by
incomplete modelling of the multiphase gas within the filaments, but
we observe almost identical variation with single phase and cooling
flow (below) models. All models assume that the gas at each
temperature has the same metallicity, though the cooler gas will
dominate in the measurement of the metallicity.

\begin{figure}
  \includegraphics[width=\columnwidth]{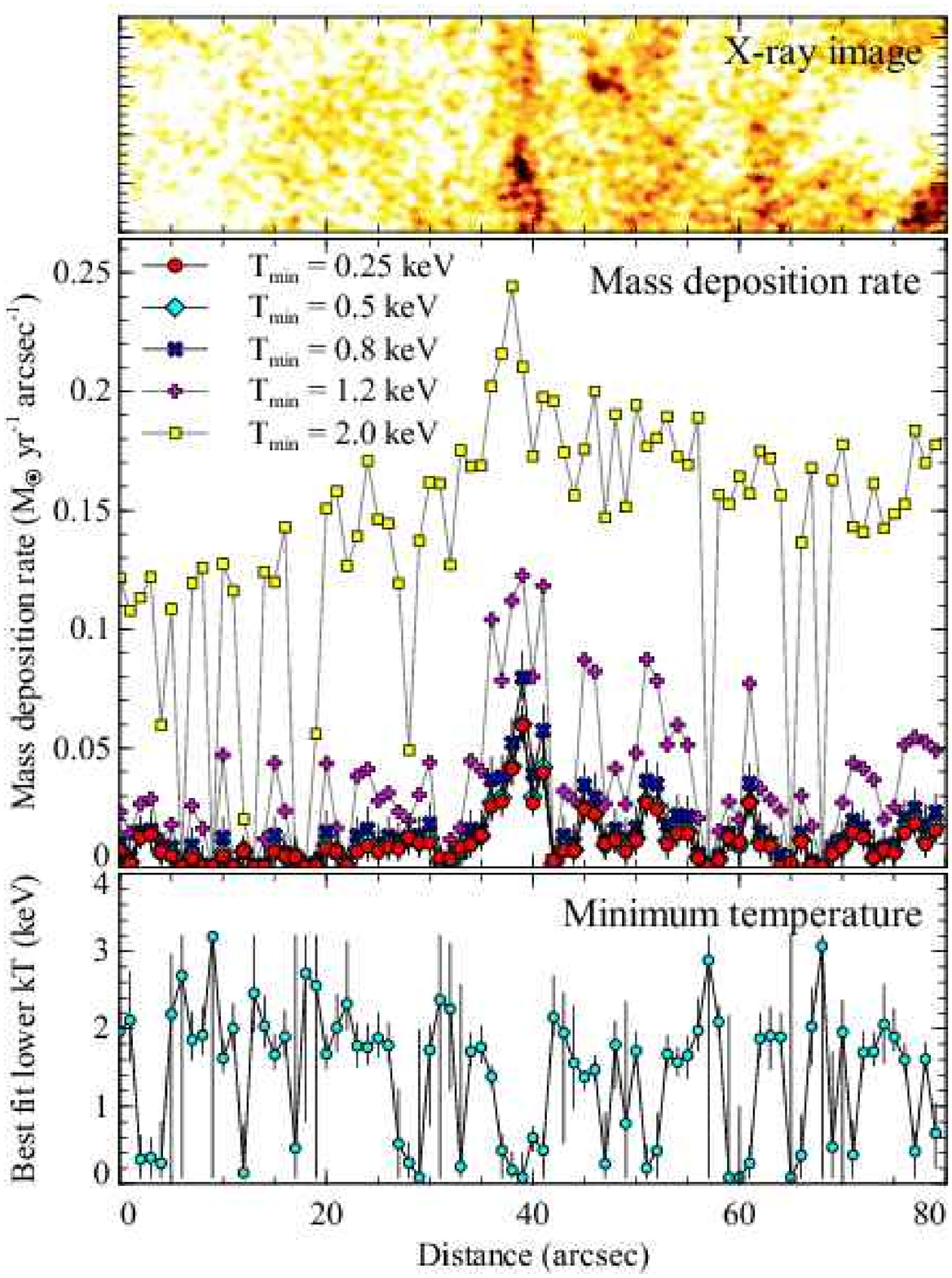}
  \caption{Results of fitting a cooling flow model to spectra
    extracted across the filaments in Fig.~\ref{fig:filamentregions}.
    The top panel shows an image of the region. In the second is
    plotted the mass deposition rates per bin obtained by fixing the
    lower temperature of the cooling flow models to certain values.
    The bottom panel shows the best fitting lower temperature if it is
    allowed to be free.}
  \label{fig:filamentcflow}
\end{figure}

\subsection{Cooling flow model}
Given the range in the temperature in the filaments, we try fitting a
cooling flow model to the spectra, as the gas may be cooling within
the filament. We used the isobaric \textsc{mkcflow} model, which
models cooling between two temperatures at a certain metallicity and
gives the normalisation as a mass deposition rate, plus a
\textsc{mekal} single phase thermal component.  The upper temperature
of the cooling flow component was tied to the temperature of the
thermal component, and the metallicities were also tied. Both
components were absorbed by a \textsc{phabs} absorber.
Fig.~\ref{fig:filamentcflow} shows the mass deposition rates obtained
with fixed minimum temperatures. The 0.25~keV rate is equivalent to a
traditional cooling flow where the gas cools out of the X-ray band,
while the others are truncated cooling flows. Also plotted is the best
fitting minimum temperature if it is allowed to vary.

The results show that within the filaments, the model adds increasing
amounts of the cooling flow component, indicating a range of
temperatures. Using a cooling flow component with a 0.25 keV minimum
temperature, the emission measure of the single phase thermal
component varies relatively smoothly over the region. There is less
apparent cooling as the minimum temperature decreases (as with the
results of \citealt{Peterson03}), until below $\sim 0.8 \keV$ where
the results are approximately consistent. If the gas is actually
cooling in these filaments, these results indicate a rate of $\sim
0.25 \Msunpyr$ in the central filament in this small region examined.
All of the $\sim 0.5 \keV$ X-ray emitting gas observed in the Perseus
cluster appears to be associated with the H$\alpha$ emitting filaments
or is close to the nucleus \citep{FabianPer06}. The lower gas
temperature measured here is consistent with zero in the centres of
the filaments, but appears to increase by a few arcsec. We note that
the best fitting absorbing column density is almost identical to the
values from the multiphase model in Fig.~\ref{fig:filamentmulti}, so
no additional absorption is required.

\begin{figure}
  \includegraphics[width=\columnwidth]{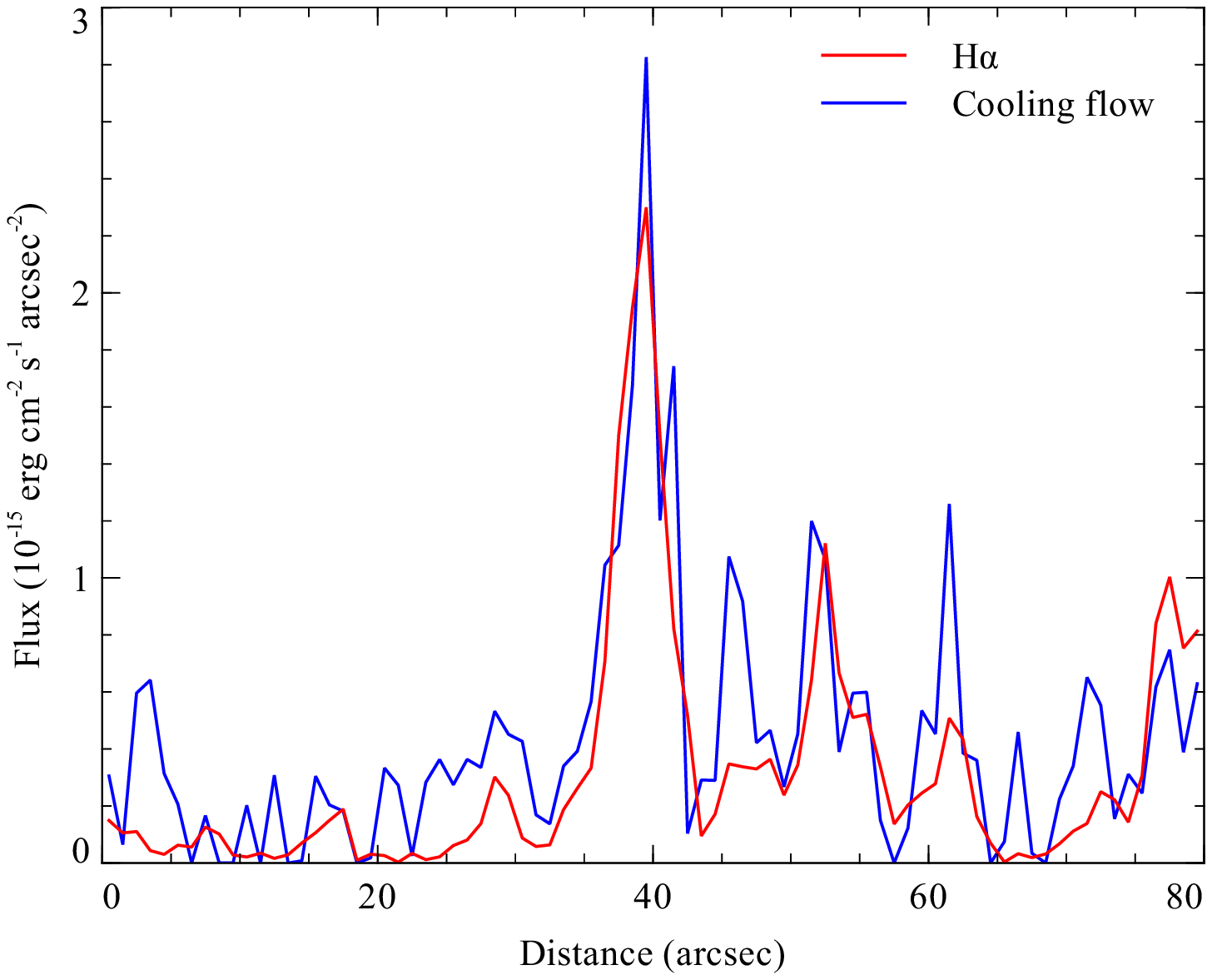}
  \caption{Comparison of H$\alpha$ surface brightness flux profile
    across the filaments (using data from \citealt{Conselice01})
    versus unabsorbed 0.001 to 30~keV surface brightness flux from the
    cooling flow model cooling down to 0.25~keV. Cooling to
    0.0808~keV, the minimum available, increases the results by about
    5~per~cent.  Both of these fluxes are in the same units.}
  \label{fig:filament_cflowflux}
\end{figure}

In Fig.~\ref{fig:filament_cflowflux} we compare the flux from the
cooling flow component cooling down to 0.25~keV (above) against the
flux in the continuum-subtracted H$\alpha$ waveband from the data from
\cite{Conselice01} (using the fluxes given in several regions to
calibrate the count to flux scale). The H$\alpha$ filter used had a
central wavelength of 6690~A and a FWHM of 77~A, and the fluxes have
been corrected for Galactic extinction, and includes N~\textsc{ii}
emission. The plot shows that the X-ray flux of the filaments,
assuming a cooling flow model, is very close to the
H$\alpha$+N~[\textsc{ii}] flux for most of the filaments. The flux from
the multitemperature model (Fig.~\ref{fig:filamentmulti}) is somewhat
similar to that from the cooling flow model. If the 2, 4 and 8~keV
components are ignored, the flux is smaller by a factor of $\sim 4$
than the cooling flow flux, but including the 2~keV flux boosts this
to well above this value.

\subsection{Filament geometry}
If the gas in a filament is in pressure equilibrium with its
surroundings, the pressure is known and thus is the temperature of
the gas in the filament, so we can estimate the volume of the emitting
region.

We fitted a simple absorbed two-temperature \textsc{mekal} spectral
model to the $1\times27.3$ arcsec bin at 38~arcsec offset where the
filaments are strongest (see Fig.~\ref{fig:filamentmulti}). The best
fitting temperatures from the spectral fit are $0.69 \pm 0.05$ and
$3.36 \pm 0.09$~keV, corresponding to the temperature of the filament
and its surroundings, respectively, under the assumption that the gas
has a single temperature in the filament. Taking the emission measure
of the filament component and an electron pressure of $0.128 \keV
\cm^{-3}$ (from the deprojected values in \citealt{SandersPer04},
Fig.~19), then the volume of the emitting region is $1.0 \times
10^{63} \cm^3$. The extracted region on the sky is $0.37 \times 10.2
\kpc$. If we assume the filament is a cuboid with the facing side of
the dimensions given, we calculate a depth of 9~pc.  If the filament
is a single long and thin tube with length 10.2~kpc, it would have
dimensions of $\sim 60 \pc$ across.

The lack of depth shows the filament is unlikely to be a sheet viewed
from the side, and the disparity with the other dimensions suggests
that the filament is in fact made up of small unresolved knots of
X-ray emission. If there were $10^3$ such blobs, they would have
dimensions of $\sim 30$~pc.

\subsection{Flux emitted from the filament}
Taking the volume above for the filament in the 38~arcsec bin and
assuming pressure equilibrium implies there are $\sim 1.9 \times
10^{62}$ electrons in the cooler filamentary component. The difference
in temperature of the cooling X-ray emitting gas from the surrounding
gas is $\sim 2.7 \keV$. This implies there would be $\sim 2.5 \times
10^{54} \erg$ released if the X-ray emitting filament was to cool out
of the surrounding intracluster medium (assuming twice the number of
particles as electrons and $3/2 \: kT$ energy per particle, which does
not include any work done by the surroundings on the cooling
filament).

The luminosity of the bin in the H$\alpha$ waveband is $6 \times
10^{40} \ergps$. This can be multiplied by a factor of 20 to account
for other line emission, for example Ly$\alpha$. Therefore if the line
emission is powered by cooling out of the intracluster medium, the
timescale for this process is $\sim 7 \times 10^4$~yr.  The dynamical
timescale for a 30~pc region, assuming a sound speed of $300 \kmps$,
would be $\sim 10^5 \yr$, so it is approximately possible for the
cooling medium to be replenished.

Conduction of heat from the surrounding intracluster medium may also
be able to power the filaments. For a 2~keV plasma, the
\cite{Spitzer62} conductivity is $\sim 9 \times 10^{11} \ergps
\cm^{-1} \K^{-1}$. If we assume a geometry of a 30~pc wide tube which
is 10.2~kpc long and if the heat is being conducted between 3.4 and
0.7~keV over 30~pc, this corresponds to a heat flux of $\sim 5 \times
10^{42} \ergps$.  Such a flux is sufficient to fuel the line
emission. It however depends greatly on the assumed conductivity
(which depends on temperature to the 5/2 power) and geometry
\citep{BohringerFabian89,NipotiBinney04}. To provide the energy for
the line emission, the heat must be able to travel to presumably
smaller and cooler regions than that of the the 0.7~keV gas, which
conductivity makes it increasingly hard to do. Note that any
conduction model requires that the surrounding soft X-ray emitting gas
and the filament have a low relative velocity. If conduction is too
efficient at transporting heat the filament will evaporate, leading to
a certain critical minimum size for growth \citep{BohringerFabian89},
which depends on how much conduction is suppressed and the temperature
and density of the surrounding ICM.

\section{Hard X-ray emission}
\label{sect:hxray}
As discussed in the introduction, \cite{SandersPer04} found evidence
for a distributed hard component surrounding the core of the cluster
by fitting a multitemperature model with a 16~keV component. The total
2-10~keV luminosity of this hard component is $\sim 5 \times 10^{43}
\ergps$. Later \cite{SandersNonTherm05} used a thermal plus powerlaw
model to fit the data giving a similar flux. This hard component has
been confirmed with \emph{XMM-Newton} data (Molendi, private
communication). Assuming this emission is the result of inverse
Compton scattering of IR and CMB photons by the population of
electrons which also emit the observed radio, the magnetic field over
the core of the cluster was mapped.

Here we examine the hard flux from the cluster using this very deep
set of observations, and investigate the effect of different spectral
models. We binned the data using the contour binning algorithm with a
signal to noise of 500 ($\sim 2.5 \times 10^{5}$ counts in each bin).

\begin{figure*}
  \includegraphics[width=\textwidth]{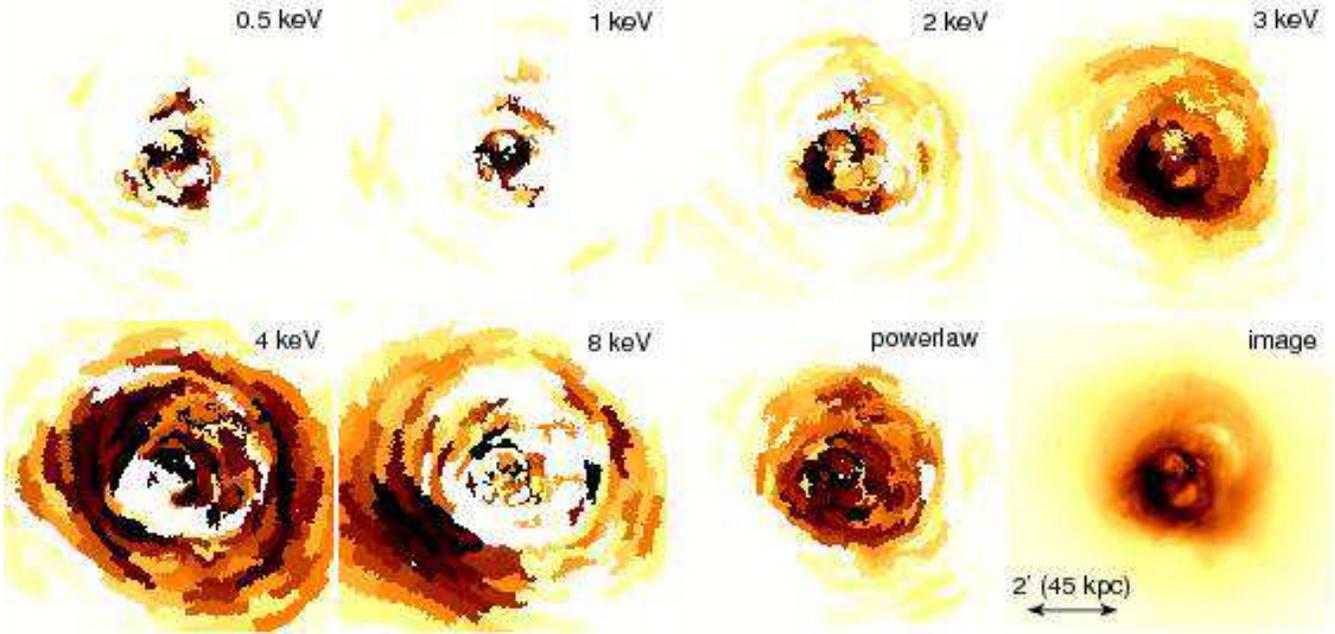}
  \caption{Results from the multitemperature plus $\Gamma=2$ powerlaw
    fits to regions containing $\sim 2.5 \times 10^{5}$ counts. The
    images show the emission measures per unit area for each of the
    components, and finally the broad band \emph{Chandra} image of the
    same region.}
  \label{fig:plawnorms}
\end{figure*}

Our first model is a multitemperature model made of different
temperature components plus a powerlaw to account for the hard
emission. This is a more complex model than that used by
\cite{SandersNonTherm05}, as we wished to account for the known cool
gas in the cluster which may affect the powerlaw signal if not
modelled correctly. Hotter gas projected from larger radii in the
cluster can also give a false signal. The data were fitted using a
model made up of \textsc{apec} thermal components at fixed
temperatures of 0.5, 1, 2, 3, 4 and 8~keV, plus a $\Gamma=2$ powerlaw,
all absorbed by a \textsc{phabs} photoelectric absorber. The
normalisations of the components were allowed to vary, the
metallicities tied to the same free value, and the absorption was
free. We used a $\Gamma=2$ powerlaw here as this is close to the best
fitting photon index of the radio emission \citep{SijbringThesis93},
and the best fitting X-ray spectral index in the core
\citep{SandersNonTherm05}. Fig.~\ref{fig:plawnorms} shows the emission
measure per unit area for each of the thermal components, the powerlaw
normalisation per unit area, and the broad band \emph{Chandra} image
of the same region.

\begin{figure}
  \includegraphics[width=\columnwidth]{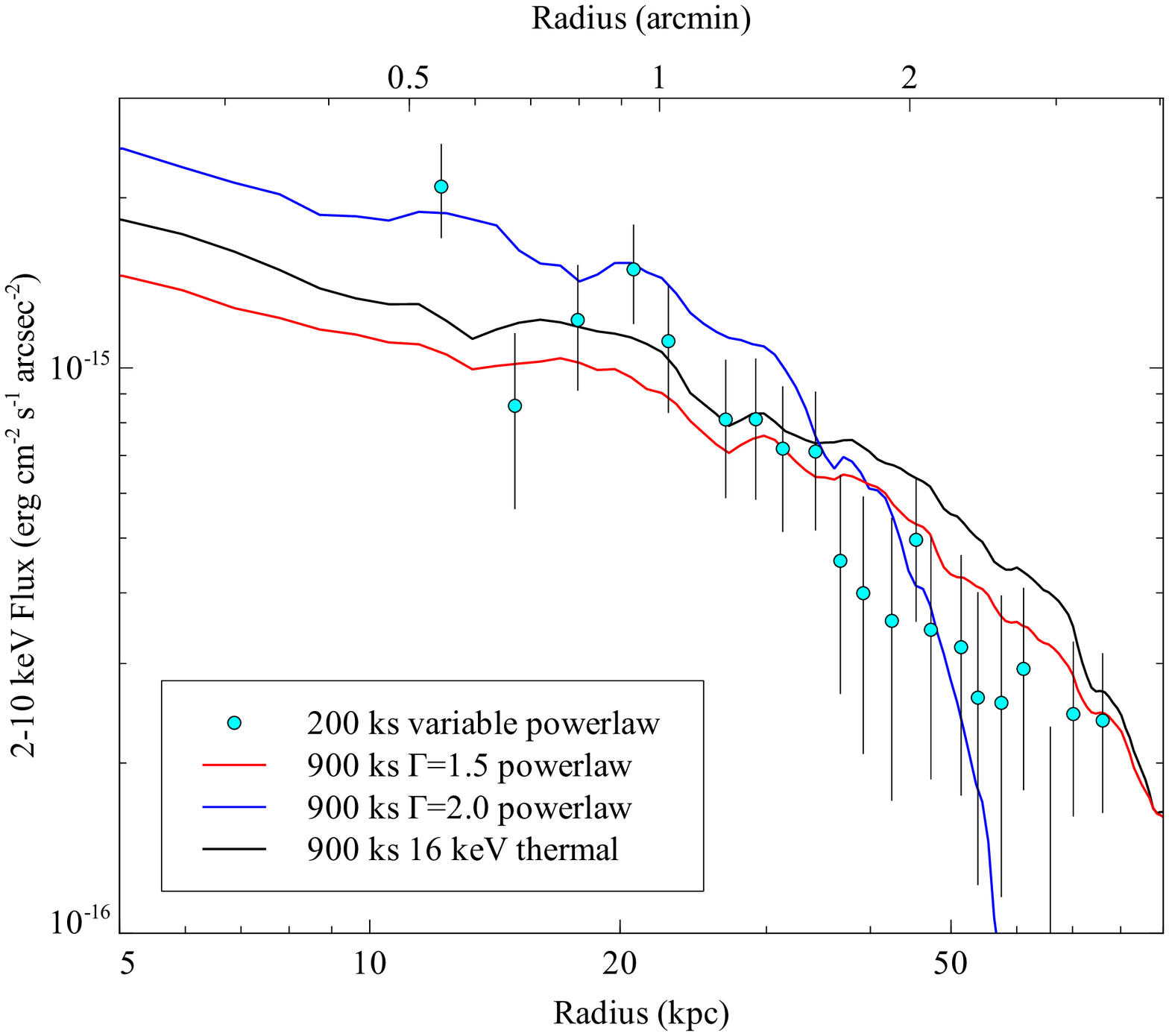}
  \caption{Flux per unit area on the sky for multicomponent models
    plus powerlaws or plus a hot 16~keV component. Also plotted are
    the background-subtracted results from an earlier analysis
    \citep{SandersNonTherm05}.}
  \label{fig:plawprofile}
\end{figure}

The thermal gas maps show similar distributions to the earlier
analyses with smaller uncertainties \citep{SandersPer04,FabianPer06}.
We plot the radial profile of the 2-10 keV flux of the powerlaw
component in Fig.~\ref{fig:plawprofile}. Also shown is the radial
profile for a $\Gamma=1.5$ powerlaw instead of the $\Gamma=2$
powerlaw, or a 16~keV thermal component, and the older results from
\cite{SandersNonTherm05} (after subtracting the estimated `background'
from hot projected thermal gas). The total emitted flux from each of
the models is fairly consistent, except at large radii where the
signal is low.  Hot thermal gas, as expected, gives similar results to
a $\Gamma=1.5$ powerlaw.  The best fitting powerlaw index looks
similar to the results in \cite{SandersNonTherm05}, with a transition
of $\Gamma \sim 2.2$ powerlaw in the centre to $\Gamma \sim 1.4$ in
the outskirts.

\begin{figure}
  \includegraphics[width=\columnwidth]{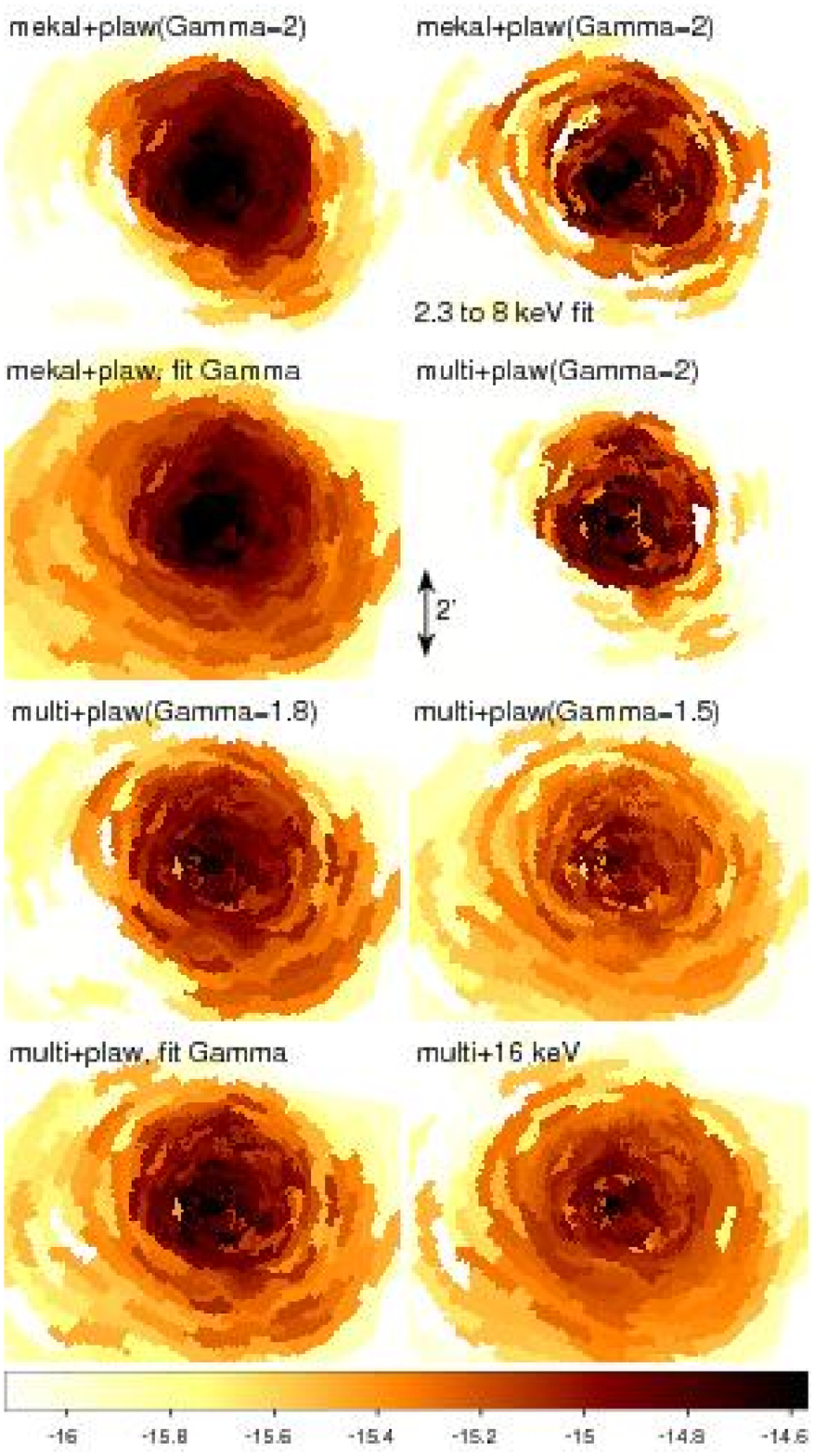}
  \caption{Fluxes for the powerlaw or hot thermal components for
    various models. The units are $\mathrm{log}_{10} \, \ergpcmsqps$,
    in the 2--10 keV band. The panels show, from left to right, top to
    bottom, (1) fitting a thermal \textsc{mekal} plus a $\Gamma=2$
    powerlaw from 0.6 to 8~keV, (2) only fitting it between 2.3 and
    8~keV, (3) allowing the $\Gamma$ to vary from 1.4 to 2.4, (4)
    multicomponent thermal \textsc{apec} model plus a $\Gamma=2$
    powerlaw, (5) using $\Gamma=1.8$, (6) using $\Gamma=1.5$, (7)
    allowing $\Gamma$ to vary between 1.4 and 2.4, and (8)
    multicomponent thermal \textsc{apec} model plus 16~keV component.}
  \label{fig:hardmaps}
\end{figure}

We map the distribution of hard flux per unit area for a variety of
different models in Fig.~\ref{fig:hardmaps}. We show the variation of
flux (top left) just using a single thermal model plus powerlaw (as in
\citealt{SandersNonTherm05}), (top right) fitting that model just in
the high energy band, and (second row left) allowing the powerlaw
index to vary. We also show the results (second row right, and third
row) from multitemperature plus powerlaw models with $\Gamma = 2, 1.8$
and 1.5, and (bottom left) fitting for the index.  Finally we show the
result (bottom right) using a multitemperature model including a 16
keV hard component.

There are problems with steep powerlaw components as they predict
significant extra flux at low X-ray energies. Such flux is not
observed, and therefore the best fitting photoelectric absorption
increases in the central regions where the powerlaw is strong. It is
possible that there is excess absorption in the central regions, but
the flux of the powerlaw is very closely correlated with the
absorption, so some physical connection between the two would be
required. Presumably the non-thermal emission process would need to be
dependent on absorbing material, which appears unlikely.

\begin{figure}
  \includegraphics[width=\columnwidth]{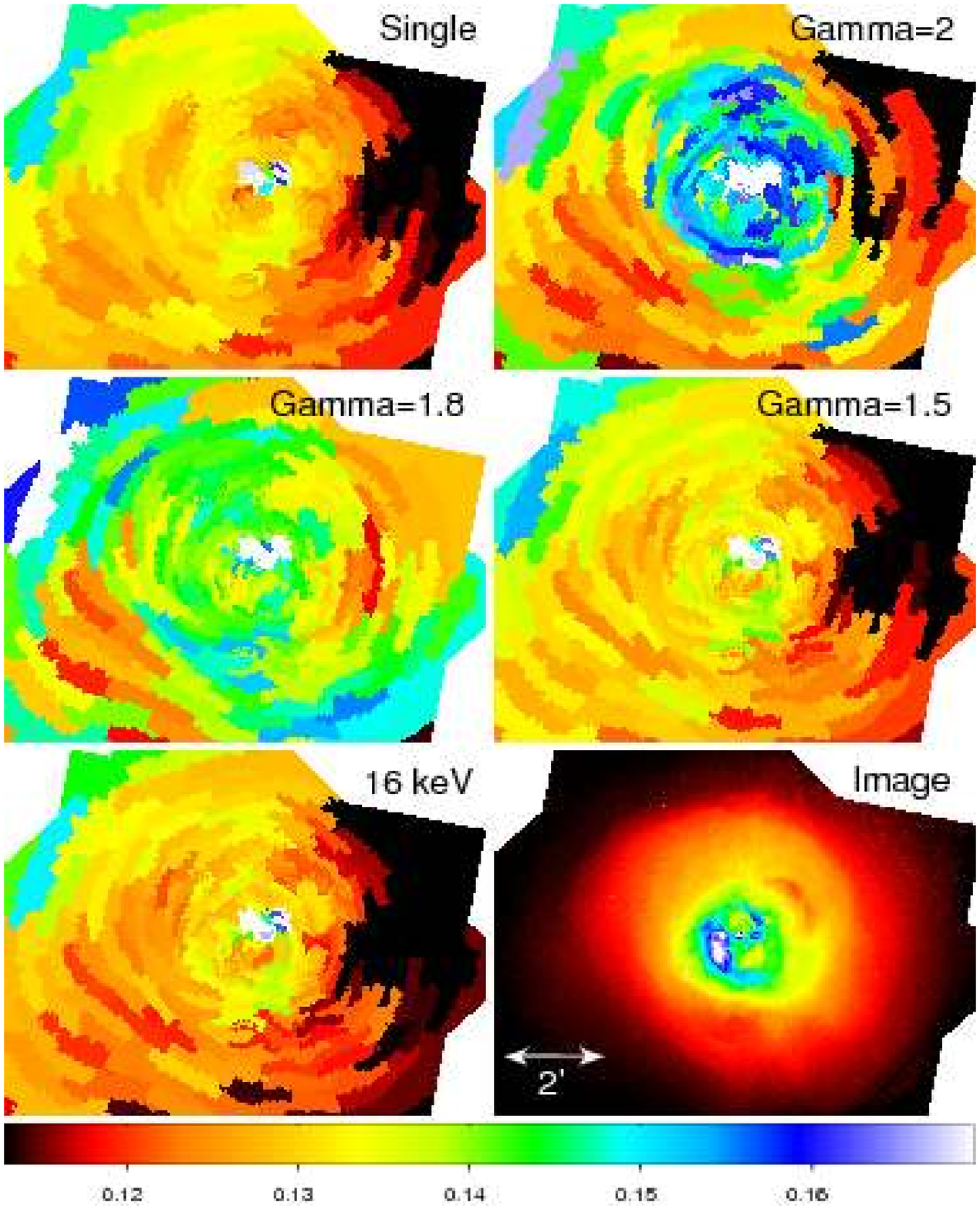}
  \caption{$N_\mathrm{H}$ column density maps in units of $10^{22}
    \psqcm$ generated by fitting different spectral models to regions
    containing a signal to noise ratio of 500. Also shown is an X-ray
    image of the same area.  `Single' shows the results of fitting a
    single \textsc{phabs} absorbed \textsc{mekal}. `Gamma=2, 1.8 and
    1.5' shows the results fitting a multitemperature model plus a
    powerlaw of the photon index given.  `16 keV' shows the results
    using a multitemperature model including a 16~keV hot thermal
    component.}
  \label{fig:nhmaps}
\end{figure}

Fig.~\ref{fig:nhmaps} shows the absorption distribution for different
models. The top-left panel shows the distribution from fitting a
single \textsc{mekal} plasma to the projected spectra. There is
obvious variation across the image. Some of this variation may be
due to the buildup of contaminant on the ACIS detector, but the
current calibration should account for this in the creation of
ancillary response matrices. Probably most of the variation is
because the cluster lies close to the Galactic plane ($b \sim
-13^\circ$). If the image is aligned to Galactic coordinates, the
variation is mostly in Galactic latitude.

Fitting using a multitemperature model plus a $\Gamma=1.5$ powerlaw or
16~keV thermal component produces absorption maps very similar to the
single temperature map. These models require no additional absorption.
The $\Gamma=1.8$ requires moderate additional absorption, and
$\Gamma=2$ produces absorption clearly correlated to the powerlaw
flux (Fig.~\ref{fig:hardmaps} centre-top row, right column).

\begin{figure}
  \includegraphics[angle=-90,width=\columnwidth]{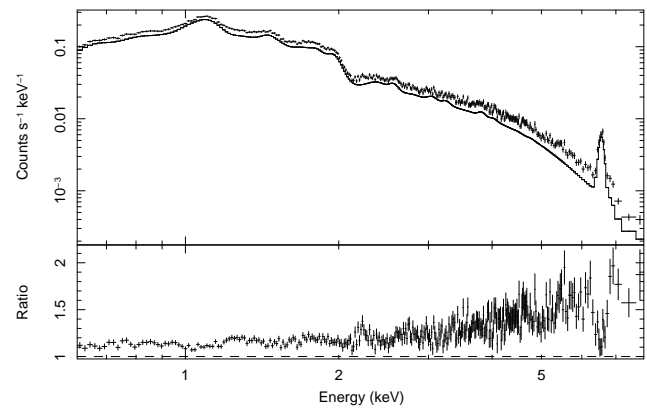}
  \caption{Spectrum showing the effect of the 16~keV component to
    spectral fit. The spectrum shown is a region around 1.8~arcmin
    north from the nucleus containing $\sim 2.5 \times 10^{5}$ counts.
    The solid line in the top panel is the model, after removing the
    16~keV component. The bottom panel is the ratio of the data to the
    model not including the hard component.}
  \label{fig:hardspec}
\end{figure}

To demonstrate the effect of the hard component on the spectral fit,
we show in Fig.~\ref{fig:hardspec} the contribution of the hard
component to the best fitting model to a spectrum from a region around
1.8~arcmin north of the nucleus. For a 16~keV component the effect is
around 10 per cent at low energies, increasing to fifty per cent at
high energies: the hard component is about one third or more flux
above 4~keV.

\subsection{Origin of the hard emission}
If we assume that the hard component is real and not an instrumental
or modelling artifact, there are two sets of possible emission
mechanisms, thermal or non-thermal. If it is non-thermal emission it
is likely to be inverse Compton emission from CMB or IR photons being
scattered by relativistic electrons in the ICM
\citep{SandersNonTherm05}. Another possible origin is hot thermal gas.
The probable origin of such material in the cluster is from a shock,
and the obvious candidate is the HVS.

\subsubsection{Thermal origin: merger of HVS}
\label{sect:hvsthermal}
From Section~\ref{sect:hvs}, the HVS lies at minimum distance from the
cluster core of 110~kpc, where the electron density of the gas is
$\sim 5 \times 10^{-3} \pcmcu$ \citep{SandersPer04}. The HVS is moving
at $3000\kmps$ relative to the main NGC\,1275 system along the line of
sight to the observer \citep{Minkowski57}. Assuming a 5~keV plasma for
the cluster, and that the galaxy is travelling along the line of
sight, the Mach number of the HVS is around 2.3.

The collision should produce a shocked Mach cone around the HVS. If
the HVS lies at its minimum distance from the cluster, using the
Rankine-Hugoniot jump conditions, the density of the post-shocked
material should be a factor of 2.1 times greater than its surroundings
($n_e \sim 10^{-2} \pcmcu$) and its temperature around 15~keV.

If the origin of the hard component is the shocked material in the
cone, and it lies at the minimum distance from the cluster, we can
estimate the depth of the cone from the emission measure of the 16~keV
component (close to the expected 15 keV temperature) using the density
above. The depth we derive from the peak of the emission is between
200-300~kpc. If this is the correct interpretation, the layer of
shocked material is extremely large. This number can be reduced
significantly if the preshocked material has a higher density, as the
depth is inversely proportional to the square of the density. At the
previous minimum distance of around $60\kpc$ from the HVS from the
core of the cluster \citep{GillmonPer04}, the minimum depth is only a
few kpc, as the density was increased by a factor of 10.  If the HVS
lies further away than our minimum distance from the cluster core, it
becomes much harder for the thermal interpretation to be a plausible
explanation. Whether a thermal origin for the hard component is
plausible depends on whether our lower limit of the distance from the
cluster to the HVS is overestimated (see Section \ref{sect:hvs}).

\subsubsection{Non-thermal origin: inverse Compton processes}
If the origin of the flux is from inverse Compton emission, there is a
large population of relativistic electrons scattering CMB or IR
photons. The bright radio emission from the mini radiohalo in the
cluster core indicates that there are relativistic electrons, but we
cannot directly observe the required $\gamma \sim 1000$ electrons by
their synchrotron emission. The powerlaw index of the inverse Compton
emission will be the same as the radio emission, if a single
population of electrons produces both. We do not know whether there is
a single population.  There could be a break in the powerlaw index, or
a cut-off in the population.

Under the assumption of a single population, we previously estimated
the magnetic field in the core of the cluster to be a 0.3-1$\mu$G
\citep{SandersNonTherm05}.

There are some potential issues with an inverse Compton explanation.
If the source electron population creates a $\Gamma \sim 2$ powerlaw,
we require significant amounts of additional absorption in the core of
the cluster (Fig.~\ref{fig:nhmaps}), as the powerlaw is significant at
low X-ray energies, unless it breaks in the X-ray band. Such
absorption is not seen on the X-ray spectrum of the nucleus using
\emph{XMM} \citep{ChurazovPer03}. The excess absorption is not
required with a flatter powerlaw index ($\Gamma \sim 1.5$).  However a
flat $\Gamma=1.5$ powerlaw will emit significant flux in the hard
($>20$~keV) band.

Hard X-rays were observed from Perseus using \emph{HEAO 1}
\citep{PriminiHeaoPer81} in the 20-50 keV band with a photon index of
around 1.9. The flux translates to a total luminosity in the 2-10 keV
band of around $1.6 \times 10^{44} \ergps$ above the thermal emission.
This component was found to be variable over a four year timescale.
\cite{NevalainenPds04} observed a flux around four times lower than
this value using the high energy PDS detector on \emph{BeppoSAX}. The
variable component must be associated with the central nucleus rather
than inverse Compton emission. The separation of the nuclear spectrum
from any hard component is difficult, particularly if the nucleus
varies on short timescales.  \cite{NevalainenPds04} concluded that the
central nucleus can account for all of the nonthermal emission
observed from Perseus using \emph{BeppoSAX}. More sensitive
measurements of the hard flux from, for example, \emph{Suzaku} are
vital to resolve this issue.  Observations with higher spatial
resolution close in time would help remove uncertainty about the
nuclear component.

The cooling time for electrons producing 10~keV electrons from inverse
Compton scattering of CMB photons is $\sim 10^9$~yr. It is therefore
likely that the spectrum could be broken at these energies. Such
breaks could reduce the need for increased absorption for steeper
powerlaw models here, or reduce the hard X-ray flux for those with
flatter spectra.

\begin{figure}
  \includegraphics[width=\columnwidth]{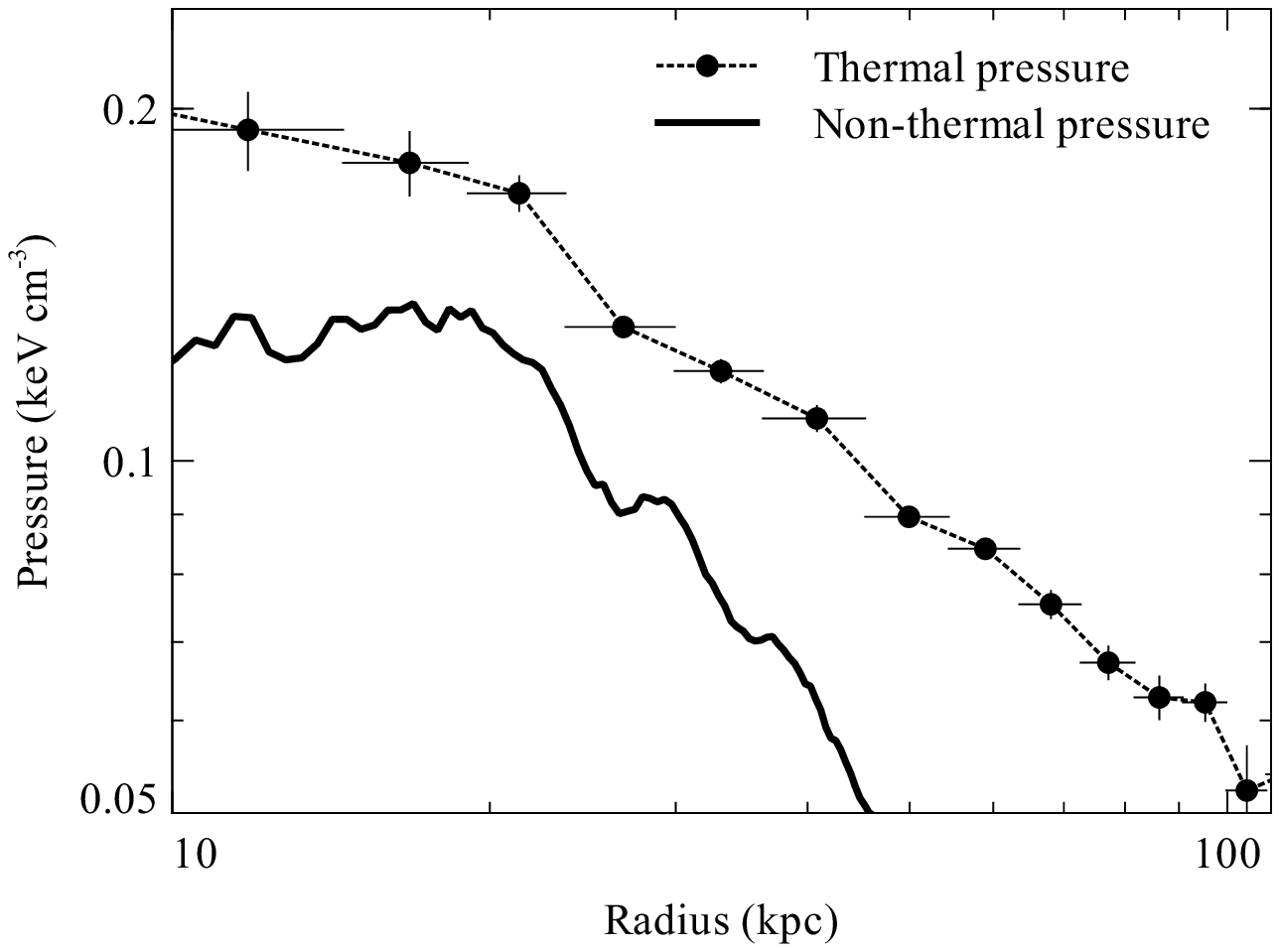}
  \caption{Inferred average nonthermal particle pressure calculated
    from the $\Gamma=1.5$ powerlaw plus multitemperature results in
    Fig.~\ref{fig:hardmaps}, assuming inverse Compton emission. Also
    plotted is the average thermal gas electron pressure from \protect
    \cite{SandersPer04}.}
  \label{fig:nonthermpress}
\end{figure}

If the emission is the result of inverse Compton emission, the
pressure of the relativistic electrons, $P$, is related to the emissivity of
the inverse Compton emission, $\mathcal{E}$, by (e.g. \citealt{Erlund06})
\begin{equation}
  P = \frac{1}{4} \frac{ \mathcal{E} \, m_\mathrm{e} \, c } { U \, \gamma
    \, \sigma_\mathrm{T} },
\end{equation}
where $U$ is the energy density of the photon field being scattered,
$\gamma$ is the Lorentz factor of the electron scattering the photon
to the observed waveband, $m_\mathrm{e}$ is the rest mass of the
electron, $\sigma_\mathrm{T}$ is the Thomson cross-section and $c$ is
the speed of light in a vacuum. If we assume that the depth of an
emitting region in the hard X-ray maps is its radius, and that the
X-ray emission is the result of scattering CMB and IR photons
\citep{SandersNonTherm05} we can estimate the nonthermal pressure as a
function of radius. We plot this in Fig.~\ref{fig:nonthermpress},
showing that the nonthermal pressure is comparable to the thermal
pressure near the centre of the cluster.

\section{Thermal content of the radio bubbles}
\label{sect:bubbles}
Earlier \emph{Chandra} data have been used to limit the amount of
thermal material material within the radio bubbles
\citep{SchmidtPer02}. This analysis is made more difficult because the
geometry of the core of the cluster is complex, and techniques
accounting for projection appear not to work generally when examining
the bubbles \citep{SandersPer04}. Here we place stringent limits on
the volume filling factor of thermal gas using this 900-ks combined
dataset using a comparative technique which depends far less on
geometry.

We fit the projected spectrum from the inside of the bubble with a
model made up of multiple temperature components at relatively low
temperatures (fixed to 0.5, 1, 2, 3, 4 and 6 keV with normalisations
varied and metallicities tied together) to account for the projected
gas, plus a component fixed to a higher temperature to test for the
existence of hot thermal gas within the bubble. We also do the fit
again with an additional $\Gamma=2$ powerlaw to account for any
possible non-thermal emission. We compare the normalisation per unit
area of the component accounting for hot thermal gas in the bubble
against that from a fit to a neighbouring region at the same radius in
the cluster. The temperature of the gas in the bubble is stepped over
a range of temperatures in the bubble and comparison regions.

The normalisation per unit area in the bubble and comparison region
can be converted into an upper limit on the difference in
normalisation per unit area between the two. We use any positive
difference between the bubble region normalisation and background,
plus twice the uncertainty on the difference (Using the positive
uncertainty on the background and the negative uncertainty on the
foreground. This is slightly pessimistic compared to symmetrising the
errors), to make a 2$\sigma$ upper limit.

Assuming a volume for the bubble, an upper limit on the density of gas
at that temperature can be calculated, assuming the gas is volume
filling. If the gas is at pressure equilibrium with its surroundings
and the pressure is known, then a limit of the volume filling fraction
of gas at that temperature can instead be calculated.

\begin{figure}
  \includegraphics[width=\columnwidth]{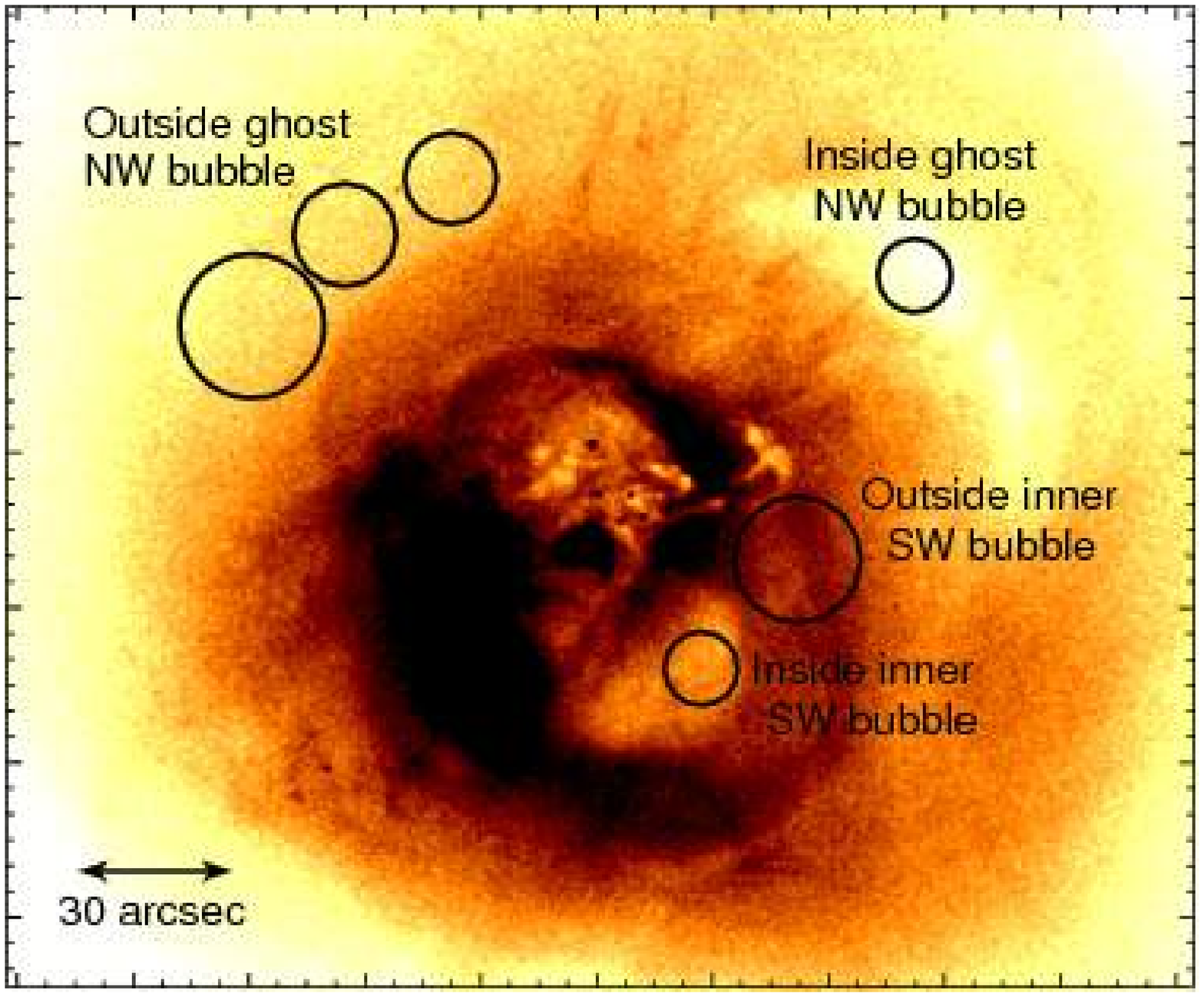}
  \caption{Regions used for examining the spectra inside and outside
    of the bubbles.}
  \label{fig:bubbleregions}
\end{figure}

We examine the inner SW bubble and ghost NW bubbles using the regions
shown in Fig.~\ref{fig:bubbleregions}. Also indicated are the regions
used for background. The inner NE bubble is obscured by the High
Velocity System, so we do not consider that. The ghost S bubble has
somewhat uncertain geometry. We try to place the regions away from any
contamination by low temperature gas (though we have tried alternative
regions with little effect), and the background regions at a similar
radius to the bubbles. We assume the bubble regions are cylindrical in
shape, with depths of 9.4~kpc and 13.4~kpc for the inner SW and ghost
NW bubbles, respectively. We take the deprojected electron pressure of
the surrounding thermal gas from the mean value of the sectors
surrounding the bubbles in figure 19 of \cite{SandersPer04}. This
leads to values of 0.195 and $0.111 \keV \pcmcu$ for the inner SW and
ghost NW bubbles, respectively.

\begin{figure}
  \includegraphics[width=0.9\columnwidth]{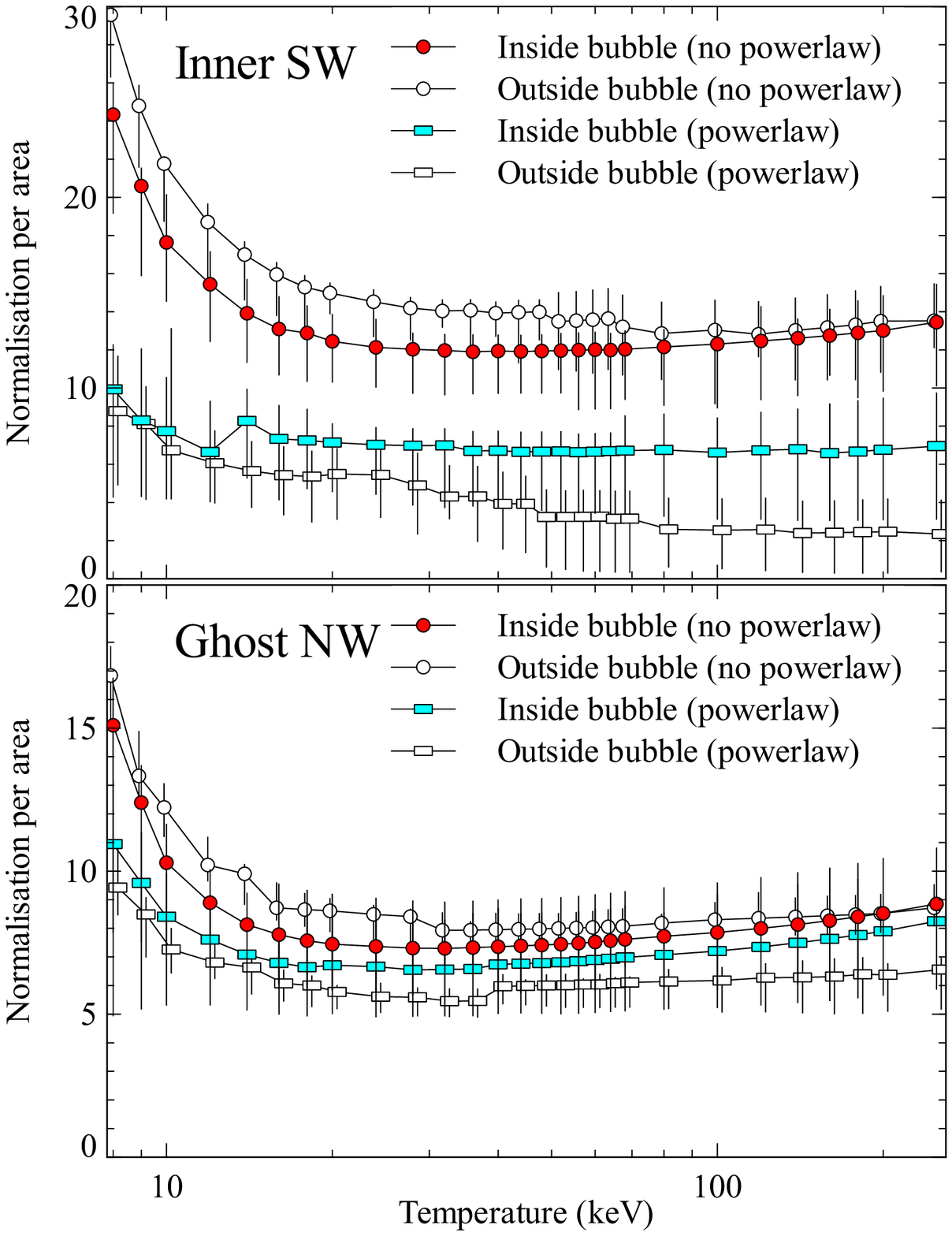}
  \caption{Hot thermal component normalisation per unit area (defined
    by the BACKSCAL header keyword) as a function of temperature
    inside the bubbles compared to outside.  The results of including
    a $\Gamma=2$ powerlaw component are also shown.}
  \label{fig:bubblenorms}
\end{figure}

\begin{figure}
  \includegraphics[width=0.9\columnwidth]{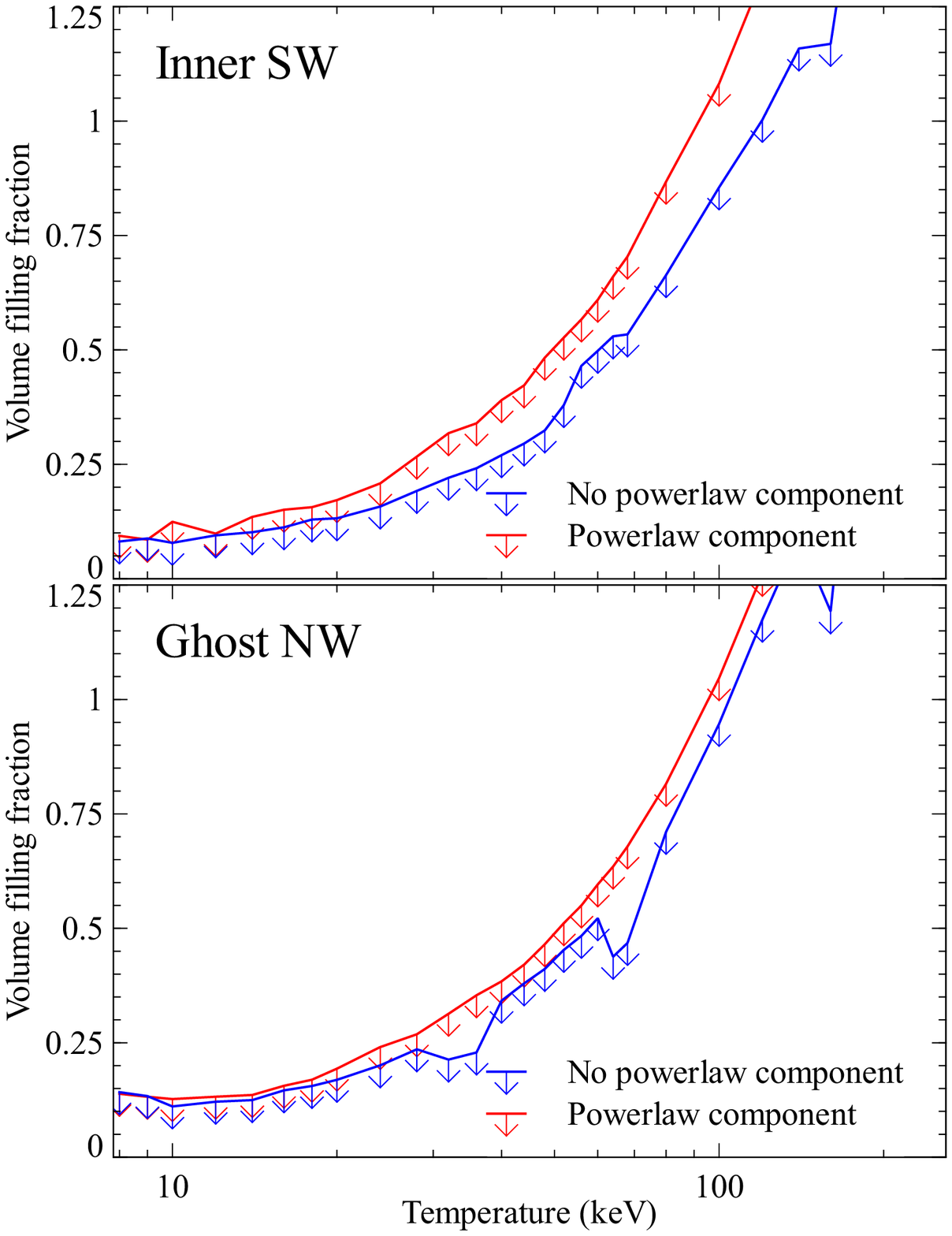}
  \caption{2$\sigma$ upper limits to the volume filling fraction of
    the bubbles. Models including a $\Gamma=2$ powerlaw component are
    indicated.}
  \label{fig:bubblevff}
\end{figure}

For the two bubbles, including and not including the $\Gamma=2$
powerlaw component, we show the normalisation (emission measure) per
unit area for the foreground and background regions in
Fig.~\ref{fig:bubblenorms}. We convert these values to upper limits on
the volume-filling fraction in Fig.~\ref{fig:bubblevff}. We find that
if there is volume-filling thermal gas within the bubbles, it must
have a temperature of less than $\sim 100 \keV$ for either bubble.

\section{Discussion}

We have investigated several issues from the Chandra data of the
Perseus cluster and now attempt to tie the interpretation of the
various phenomena more closely together. In particular we try to
understand the interplay between heating and cooling in the cluster
core, and how energy is transported and distributed  through the ICM.

\subsection{Sound waves}

Observations of Perseus and many other similar clusters show that jets
feed relativistic plasma from around the massive central black hole
into twin radio lobes. The power associated with this process is large
and comparable to that required to balance radiative cooling within
the regions of 50--100~kpc where the radiative cooling times is
3--5~Gyr \citep{Rafferty06,DunnFabian06}. In the case of the Perseus
cluster the jets are producing on average around $5\times
10^{44}-10^{45}\ergps$ in total, as estimated from the $P\mathrm{d}V$
work done over the 5-10~Myr age of the bubbles
\citep{FabianCelottiPer02,DunnFabian04}. These estimates depend on the
filling factor of the bubbles by the relativistic plasma, which from
the results given in Section \ref{sect:bubbles} could well be unity.
Therefore at least about 20--40 per cent of the bubble power goes into
sound waves. 

We expect the jets to provide power more or less continuously over the
lifetime of the cluster core (several Gyr). This is supported by the
high incidence of bubbles found in cluster cores that require heating
\citep{DunnFabian06} and by the train of ghost bubbles seen in our
Perseus cluster data \citep{FabianPer06}. Note however, that there
is probably much fluctuation in the jet power over short timescales
(the radio source has weakened in strength over the past 40~yr), but
variations over a cooling time of $10^8\yr$ are probably less than an
order of magnitude.

We now address the question of how the energy in the bubbling process,
occurring on a timescale of 10~Myr, is fed into the bulk of the core
and dissipated as heat which balances the cooling. The smooth cooling
time profiles seen in cluster cores, and the peaked metal
distributions argue for a relatively gentle, continuous (on timescales
of $10^7-10^8\yr$ or more) and distributed heat source. The inflation
of the bubbles does $P\mathrm{d}V$ work on the surroundings and thus
creates pressure waves -- sound waves -- which carry the energy
outward in a roughly isotropic manner. We discovered ripples in
surface brightness in the first 200~ks of the Perseus cluster data
\citep{FabianPer03} which we interpreted sound waves and have extended
the analysis of them here.

Fig.~\ref{fig:wavepowcuml} shows that there is considerable power in
these waves, around $2-3\times 10^{44}\ergps$ at radii of 30--70~kpc.
It is similar to the level of heat required to offset radiative
cooling within that region. The wave power drops with radius
between 30 and 100 kpc indicating that the energy is being dissipated
and confirming the idea that viscous dissipation of the sound waves is
the distributed heat source. The dissipation length is comparable to
that estimated on the basis of Spitzer-Braginski viscosity
\citep{FabianPer03}, although the process in the likely tangled
magnetic, and cosmic-ray infused, plasma may be more complex and
involve some form of bulk viscosity.

There is however a deficit of power shown by this analysis within
30~kpc.  Although it could just be variability of the central power
source (the jets), our results on the nonthermal component associated
with the radio minihalo indicates that there is a significant
nonthermal pressure from the cosmic rays there, comparable to the
thermal pressure. This would quadruple (if the nonthermal wave
pressure is the same as the thermal pressure in Equation
\ref{eqn:wavepower}) the predicted wave power in this region, bringing
it into agreement with the region between 30 and 60~kpc and meaning
that dissipation of sound waves can be the dominant heat source
balancing radiative cooling.

The increase in power seen in Fig.~\ref{fig:wavepowinst} beyond
100~kpc can be explained as associated with the power $\sim 10^8\yr$
ago being 2--3 times larger than present. This may fit in with the
presence of the long Northern optical filament \citep{Conselice01},
which if drawn out from the centre by a bubble \citep{HatchPer06} must
have been an exceptional bubble, possibly seen at 170~kpc north of the
nucleus (Section \ref{sect:wavepower}). Also the ghost bubbles seen to the South in the X-ray
pressure maps \citep{FabianPer06} may be larger and so require more
power than the present inner bubbles. If correct, the central source
thus does vary by a factor of few on timescales of $10^8\yr$.  Deeper
X-ray observations at larger radii than the current ones are needed to
test this interesting possibility further.

We conclude the discussion of sound waves by noting that typical sound
waves will be difficult to see in current data on other clusters.  The
X-ray surface brightness of the central 50~kpc of the Perseus cluster
is more than twice that of another cluster. Large ripples, interpreted
as weak shocks have been seen in the Virgo cluster \citep{FormanM8706}
and possible outer ripples may occur in A\,2199 and and 2A\,0335+096
\citep{SandersShock06}. These may just be the peaks in the
distribution of sound waves in those clusters from the more
exceptional power episodes of their central engines. Like tree rings or
ice cores for the study of geological history, ripples in cluster
cores offer the potential to track the past history of an AGN for more
than $10^8\yr$. Observations of a yet wider region are required to see
where the ripples eventually die out.  

\subsection{The nonthermal component}

We have reported on a hard X-ray component coincident with the radio
minihalo (section \ref{sect:hxray}). It is difficult to support a
thermal origin for this emission in terms of a $\sim 16\keV$ gas
associated with the HVC.  A simple energy argument against the thermal
hypothesis is to note that the crossing time of the 200~kpc diameter
inner core of the cluster at $3000\kmps$ takes nearly $7\times
10^7\yr$. The radiative cooling time of 16~keV gas with a plausible
density of $10^{-2}\pcmcu$ is $10^{10}\yr$, so the shocked gas would
radiate only 0.7 per cent of its energy. With the luminosity of this
component at $5\times 10^{43}\ergps$ we obtain a total injected energy
of $2\times 10^{61}\erg$. This is the total kinetic energy of $2\times
10^{11}\Msun$ moving at $3000\kmps$. So an implausible large mass
needs to be stripped from an apparently small galaxy in order to
explain the hard component as due to shocking by the HVC.

The hard component is therefore most readily interpreted as inverse
Compton emission from the minihalo. As discussed by
\cite{SandersNonTherm05} this means that it must be well out of
equipartition for the electrons and magnetic field with the electrons
dominating the nonthermal pressure (as also deduced for the radio
lobes, \citealt{FabianCelottiPer02}). It is possible that the radio
bubbles leak a small\footnote{Studies of bubbles in nearby clusters
  indicate they are not magnetic pressure dominated
  \citep{DunnFabian04,DunnFabian06}, so this limits the number of
  particles which can escape.} fraction of their cosmic-ray electron
content (and presumably protons or positrons for charge neutrality)
into their surroundings.  These accumulate, losing their energy
principally through inverse Compton losses on the Cosmic Microwave
Radiation.  A half-power break in the power-law spectrum is then
expected around 10~keV in the hard X-ray flux if this process has, as
expected, been continuing for more than a Gyr. This matches what is
inferred of the observed spectrum.

The presence of the nonthermal component increases the heating within
the inner 30~kpc if $\delta P$ is raised proportionately, and also
means that some direct collisional heating of gas is
possible. \cite{RuszkowskiCosmicRay07} have since investigated this
idea in more detail.

\subsection{The distribution of metals}

The enhanced metallicity in the core of the Perseus cluster shows a
spatial distribution which extends to the S along the axis of old
bubbles, as expected if the bubbles push and drag the gas around
\citep{Roediger07}. What is particularly interesting here is
evidence that the distribution is clumpy and also that the central
drop is a real effect.

The clumpiness and especially the sharp edges of at least one clump
(Fig.~\ref{fig:blob_profile}) likely require a magnetic field
configured to prevent dispersal. The central metallicity drop is not
easily explained as just due to some outer gas falling in to replace
inner gas dragged out, since it has a significantly lower entropy. It
could be accounted for if the metals are highly inhomogeneous on a
small scale, with the higher metallicity gas which has the shorter
cooling time cooling out \citep{MorrisFabian03}.

An interesting possibility raised by the high nonthermal pressure near
the centre is that the gas is made buoyant by the cosmic rays
\citep{Chandran04,Chandran05}.  In this picture the simple entropy
inferred from just the gas temperature and density are insufficient to
determine its behaviour.  The gas cosmic-ray mixture becomes
convectively unstable where the cosmic-ray pressure begins to drop
steeply outward, leading to the gas overturning.  The pressure drop is
inferred to occur at about 30~kpc (Fig.~\ref{fig:nonthermpress}) and
is about the radius where the metallicity peaks
(Fig.~\ref{fig:Zmap}). Such a turnover of gas may happen sporadically
when the cosmic-ray density has built up for some time.

\subsection{The optical filaments}

Finally we briefly discuss the H$\alpha$ filaments. These radiate most
of their energy at Ly$\alpha$ (see \citealt{FabianNulsen84} for an
image) and are mostly composed of molecular gas at a few thousand down
to 50~K \citep{Hatch05,Johnstone07,Salome06}. We have shown that they
are surrounded and mixed with soft X-ray emitting plasma at 0.5--1~keV
temperature. In section \ref{sect:proffilament} we consider one
filament in detail. The H$\alpha$ luminosity of its peak is about one
per cent of the total measured by \cite{Heckman89}, so scaling the
X-ray inferred mass cooling rate there of $0.06\Msunpyr$ we obtain a
total mass cooling rate into filaments of $5\Msunpyr$.

This value assumes however that the X-ray emitting gas loses its
energy solely by radiating X-rays. If in addition it loses energy by
conduction, mixing or other means, with the cold gas in the filament
and thereby powers the Ly$\alpha$ emission then we need to scale the
above rate by a factor of 20 (the expected Ly$\alpha$/H$\alpha$ ratio
for recombination) to obtain a total mass cooling rate of about
$100\Msunpyr$. This value agrees of course with that obtained by just
taking the total H$\alpha$ luminosity and assuming it is obtained from
the thermal energy of the hot gas. It is about one third of the total
inferred mass cooling rate obtained if there is no heating. We note
that radiation at other wavelengths such as O~\textsc{vi} emission
\citep{Bregman06} can increase this fraction.

This implies that a high non-radiative mass cooling rate is possible
in the Perseus cluster. By extension, it suggest that this happens in
most cool-core clusters (which generally also have optical filaments;
see \citealt{CrawfordBCS99}). It can account for \emph{part} of the
lack of cool X-ray emitting gas in such cluster cores
\citep{FabianCFlow02}.

A cooling rate of $100\Msunpyr$ for 5~Gyr gives a total of $5\times
10^{11}\Msun$ of cold gas. This is about 10 times more than is
inferred from CO measurements \citep{Salome06}. Star formation is
another possible sink. NGC\,1275 has long been known to have an A-star
spectrum and excess blue light. UV imaging of NGC\,1275 by
\cite{Smith92} shows a lack of stars above $5\Msun$, which means that
any continuous star formation (which would have been only
$20\Msunpyr$) must have ended about 50~Myr ago. Burst models of star
formation at rates up to many $100\Msunpyr$ 100~Myr or so ago are
consistent with the UV data \citep{Smith92}. A picture in which gas
accumulates through filaments and is then converted into stars
sporadically on a 100~Myr timescale appears possible.
 
If however much of the power in the filaments is due to sources other
than the hot gas, cosmic rays or kinetic motions for example, then the
above star formation estimate is an upper limit.

\subsection{Summary}

We have quantified the properties of the X-ray surface brightness
ripples found in the core of the Perseus cluster and, assuming that
they are due to sound waves, have determined the power propagated as
sound waves. The power found in this way is sufficient to balance
radiative cooling within the inner 70~kpc, provided that it is
dissipated as heat over this lengthscale.  This provides the
quasi-isotropic, relatively gentle, heating mechanism required to
prevent a full cooling flow developing. Ultimately the power is
derived from the jets emitted by the central black hole. 

A hard X-ray component is confirmed and argued to be plausibly due to
inverse Compton scattering by cosmic-ray electrons in the radio
minihalo. The electrons may have leaked out of the radio lobes of 3C84
and now have a pressure comparable to the thermal pressure of the hot
gas in the innermost 30~kpc. The lobes themselves appear to be devoid
of any thermal gas unless its temperature is very high (50--100~keV).
The cosmic-ray electrons are important in enhancing the heating and
possibly also in changing the convective stability of the central
30~kpc.

The X-ray data provide insight on the history of the past 100~Myr of
activity of the nucleus of NGC\,1275. There are hints from the large
ripples beyond 100~kpc, and from the the large Northern filament and
the presence of many A stars, of a higher level of activity before
that, a few 100~Myr ago. This can be tested by deep, high spatial
resolution, observations of a wider region than covered out so far.

We infer that a close balance between heating and cooling is
established in the core of the Perseus cluster over the past few
100~Myr. The average heating rate is, and has been, close to the
radiative cooling rate, although there can be variations by a factor
of a few on longer timescales. The primary energy source is the
central black hole and jets; the energy is distributed by the sound
waves generated by the inflation of the lobes. We suspect that this
process is common to most cool core clusters and groups and is the
mechanism by which heating of the cool core occurs. It will however be
difficult to verify observationally in those other objects since the
X-ray surface brightness is so much lower. Only the extreme peaks in
the distribution of activity will generally be detectable.

\section*{Acknowledgements}
ACF acknowledges The Royal Society for support. We thank the
\emph{Chandra} team for enabling the superb images of the Perseus
cluster to be obtained.

\appendix

\section{A Direct Spectral Deprojection Method}
\label{appendix:deproj}
\subsection{Problems with existing spectral deprojection methods}
Results from fitting cluster data using the \textsc{projct} model in
\textsc{xspec} can be misleading.  \textsc{projct} is a model to fit
spectra from several annuli simultaneously, to account for projection.
There are one or more components per deprojected annulus, each with
parameters (e.g.  temperature, metallicity). The projected sum of the
components along line of sights (with appropriate geometric factors)
are fitted against each of the spectra simultaneously.  Often the
resulting deprojected profiles (e.g. temperature) oscillate between
values separated by several times the uncertainties on the values.
This oscillation can disappear if different sized annuli are used.
Sometimes halving the annulus width can halve the oscillation period,
indicating they are not physical changes on the sky.

The oscillation can be alleviated by fitting the shells sequentially
from the outside, freezing the parameters of components in outer
shells before fitting spectra from shells inside them (see e.g.
\citealt{SandersPer04}).  This helps to solve the issue where poorly
modelled spectra near the centre can affect the results in outer
annuli (The standard way to fit the data is simultaneous. All the
spectra are used to calculate each point, even though interior shells
and not projected in front of outer shells.) The difficulty with this
method is that uncertainties calculated on parameters to the model are
underestimated.  They do not include the uncertainties on outer
shells.

The outside-first fitting procedure does not fix every oscillating
profile. Numerical experiments, when clusters are simulated and fit
with \textsc{projct} (R.~Johnstone, private communication), show that
assuming the incorrect geometry when trying to account for projection
does not produce oscillating profiles.  Something which does create
oscillating profiles are shells which contain more than one spectral
component (e.g. several temperatures).  It appears that
\textsc{projct} tends to account for one of the components in one of
the annulus fit results, and another in a different shell. By assuming
that a spectral model is a good fit to the data, a very misleading
result is produced. Other deprojection methods which assume a spectral
model to do the deprojection will have similar issues.

\subsection{Direct spectral deprojection}
We describe here a method to create `deprojected spectra', which
appears to alleviate some of the issues found using \textsc{projct}.
It is a model independent approach, assuming only spherical
geometry (at present).

The routine takes spectra extracted from annuli in a sector, and their
blank-sky background equivalents.  From each of these foreground count
rate spectra we subtract the respective blank sky background spectrum.
Taking the outer spectrum, we assume that it was emitted from part of
a spherical shell, and calculate the (count rate) spectrum per unit
volume. This is then scaled by the volume projected onto the next
innermost shell, and subtracted from the (count rate) spectrum from
that annulus. After subtraction we calculate a spectrum per unit
volume for the next innermost shell.  We move inwards shell-by-shell,
subtracting from each the calculated contributions from outer shells.
This yields a set of deprojected spectra which are then directly
fitted by spectral models.

To calculate the uncertainties in the count rates in the spectral
channels in each spectrum we used a Monte Carlo technique. Firstly
each of the input foreground and background spectra are binned using
the same spectral binning, using a large number of counts per spectra
channel (we used 100 in this work) so that Gaussian errors can be
assumed in each spectral bin. We repeat the deprojection process
5000 times, creating new input foreground and background spectra by
simulating spectra drawn from Gaussian distributions based on the
initial spectra and their uncertainties. The output spectra are the
median output spectra from this process, and the 15.85 and 84.15
percentile spectra were used to calculate the 1$\sigma$ errors on the
count rates in each channel.

This technique assumes that the response of the detector does not
change significantly over the detector. This is the case for the
ACIS-S3 detector on \emph{Chandra} used here. It also assumes that the
effective area does not change significantly, as we do not account for
the variation of the ancillary response. This could be incorporated
into this method, but we have not implemented this yet.

\bibliographystyle{mnras}
\bibliography{refs}

\clearpage

\end{document}